\documentclass[prb,aps,amsmath,twocolumn,amssymb,floatfixng,showpacs,
superscriptaddress]{revtex4}
\usepackage[dvips]{graphics}
\usepackage{hyperref}
\usepackage{color}
\definecolor{dred}{rgb}{0.6,0,0}
\pdfoutput=1

\begin{document}

\title{Magneto-transport in closed and open mesoscopic loop structures: 
A review}

\author{Paramita Dutta}

\email{paramita.dutta@saha.ac.in}

\affiliation{Condensed Matter Physics Division, Saha Institute of 
Nuclear Physics, Sector-I, Block-AF, Bidhannagar, Kolkata-700 064, India} 

\author{Santanu K. Maiti}

\email{santanu.maiti@isical.ac.in}

\affiliation{Physics and Applied Mathematical Unit, Indian Statistical
Institute, 203 Barrackpore Trunk Road, Kolkata-700 108, India}

\begin{abstract}

Magneto-transport properties in closed and open loop structures are 
carefully reviewed within a tight-binding formalism. A novel mesoscopic 
phenomenon where a non-vanishing current is observed in a conducting 
loop upon the application of an Aharonov-Bohm flux $\phi$ and we
explore its behavior in the aspects of quantum phase coherence, 
electron-electron correlation and disorder. The essential results are
analyzed in three different parts. First, we examine the behavior
of persistent current in different branches of a zigzag carbon nanotube 
within a Hartree-Fock mean field approach using the second quantized 
formulation. The phase reversal of persistent current in several
branches as a function of Hubbard interaction is found to exhibit 
interesting patterns. Our numerical results suggest a filling-dependent
metal-insulator transition in a zigzag carbon nanotube. Next, we 
address the behavior of persistent current in an ordered-disordered
separated nanotube keeping in mind a possible experimental realization
of shell-doped nanowire which can provide a strange electronic behavior
rather than uniformly doped nanowires. Finally, we focus our attention 
on the behavior of persistent current in an open loop geometry where
we clamp an ordered binary alloy ring between two ideal semi-infinite
electrodes to make an electrode-ring-electrode bridge. From our 
investigation we propose that under suitable choices of the parameter 
values the system can act as a $p$-type or an $n$-type semiconductor.

\end{abstract}


\maketitle

\section{Introduction}

The abundant progress in nano-science and technology has stimulated us to 
fabricate different artificial nano-structures over the last few decades
whose dimensions are comparable to and even smaller than the mean free 
paths or wavelengths of electrons such that they can move through the
samples without randomizing their phase memories~\cite{datta1}. The 
manifestation of phase coherence length is one of the most striking 
aspects in the mesoscopic regime which can be obtained by lowering the 
temperature of the samples below sub-Kelvin temperatures since the 
scatterings due to electron-electron and electron-phonon interactions 
are highly suppressed in this temperature regime. Such experimentally 
accessible low-temperatures also make the system energy levels behave 
like discrete states which play a pivotal role in appearing several 
quantum-mechanical phenomena. Therefore, a mesoscopic system which can be
modeled as a phase-coherent elastic scatterer~\cite{jaya1} provides us 
the opportunity to explore various novel quantum-mechanical effects beyond 
the atomic realm~\cite{kramer}. The experimental realization of different 
quantum-mechanical incidents like, universal conductance fluctuations, 
non-local current-voltage relationship, new Onsager reciprocity relations, 
Coulomb blockade in micro-tunnel junctions~\cite{altshular,beenakker,
washburn,jaya2,jaya3}, Anderson localization~\cite{anderson}, quantum 
Hall-effect~\cite{klitzing}, Aharonov-Bohm (AB) oscillations~\cite{aharonov,
mooij,hod,nich}, to name a few, has raised the popularity of the mesoscopic 
world among the scientists and engineers. Another reason behind this 
popularity can be attributed to the tailor-made geometries which look 
very simple but have high potential from the application perspective. 
In addition, several other fluctuation patterns of the conductance are 
easily reproducible simply by tuning external parameters like, magnetic 
field, electric field, Fermi level, etc. 

The existence of dissipationless current, the so-called persistent current, 
in a mesoscopic normal metal ring pierced by an AB flux $\phi$ is one of
such remarkable effects which reveals the importance of phase coherence of 
electronic wave functions in low-dimensional quantum systems. The 
alluring question of persistent current in normal-metal rings threaded
an AB flux was first explored during 1960's~\cite{yang}. Later, in $1983$, 
B\"{u}ttiker {\em et al.} have revived the interest regarding this 
topic~\cite{butt1}. They have predicted that a small isolated normal metal 
ring penetrated by a slowly varying magnetic flux carries an equilibrium 
current which does not decay and circulates within the sample. But its 
experimental realization was somewhat difficult because at that time it 
was a challenging task to confine magnetic flux in such a small region like 
nanoscale ring. However, a few years later Levy {\em et al.}~\cite{levy} 
have given first experimental evidence of persistent current in a mesoscopic 
metallic ring. They have observed the oscillations with period $\phi_0/2$ 
($\phi_0=ch/e$, the elementary flux-quantum) while measuring persistent 
current in an ensemble of $10^7$ independent Cu rings. Similar oscillations 
with period $\phi_0/2$ have also been reported not only for an ensemble of 
disconnected $10^5$ Ag rings~\cite{mailly1} but also for an array of $10^5$ 
isolated GaAs-AlGaAs rings~\cite{mailly2}. Later, many other 
theoretical~\cite{gefen,ambe,schm1,schm2,san1,new1,new2,new3,new4,new5} 
as well as 
experimental~\cite{mailly3,jari,deb,blu} attempts have been done to explore 
the actual mechanisms of persistent current in single-channel rings and 
multi-channel cylinders. However, a controversy still persists among the
experimental observations and theoretical estimates of persistent current
amplitudes. All the experimentally measured current amplitudes were found 
to be one and two orders of magnitude larger than the theoretically predicted 
results, except in the case of nearly ballistic GaAs-AlGaAs 
rings~\cite{mailly3} where $\phi_0$-periodic persistent currents have been 
observed with amplitude $I_0 \sim e v_F/L$ ($v_F$ and $L$ are the Fermi 
velocity and circumference of the ring, respectively), which is very close 
to the value obtained from the free electron theory at absolute zero 
temperature ($T=0\,$K). In a recent work, Bluhm {\em et al.}~\cite{blu} 
have examined magnetic response of $33$ individual cold mesoscopic gold 
rings, considering one ring at a time, using a scanning SQUID technique. 
Their results agree well with the theoretically estimated value~\cite{gefen} 
in a single ballistic ring~\cite{mailly3} and an ensemble of $16$ nearly 
ballistic rings~\cite{mailly4}. But, the amplitudes of persistent current 
in single-isolated-diffusive gold rings~\cite{chand} are still two orders of 
magnitude larger than the theoretical calculations. This discrepancy 
initiated intense theoretical activity, and it is generally believed that 
the electron-electron correlation plays an important role in the disordered 
diffusive rings~\cite{abraham,bouzerar,shastry}. An explanation based on 
the perturbative calculation in presence of interaction and disorder has 
been proposed and it seems to give a quantitative estimate closer to the 
experimental results, but still it is less than the measured currents by 
an order of magnitude, and the interaction parameter used in the theory 
is not well understood physically. 

The behavior of persistent current in an isolated loop geometry enclosing
a magnetic flux is highly sensitive to the location of Fermi level of the 
system where the role of magnetic flux is essentially to destroy the 
time-reversal symmetry and as a result, the degeneracies among the 
current-carrying states get removed. Depending on the Fermi level and the 
direction of the magnetic flux current flows in either direction revealing 
diamagnetic or paramagnetic nature. Since the discovery persistent current 
have been studying in different kinds of mesoscopic loop structures. Among 
them single channel mesoscopic rings~\cite{gefen,san2} were mainly in the 
focus of attention in spite of having topological simplicity they carry 
a high potential from the application point of view. Not only single isolated
rings but also array of such rings have been used to study the nature of 
persistent current. A very few works on multi-channel closed loop systems 
have been discussed~\cite{gefen2,wohlman1,wohlman2}. In case of 
multi-channel mesoscopic cylinders it has been noticed that the typical 
single-level current gets reduced with increasing number of conducting 
channels $M$. The correlations in the energy spectrum governs that the 
ratio of the total and single-level currents is proportional to 
$\sqrt M$~\cite{gefen2}.

Similar to persistent current in such isolated closed loop geometries, 
the non-decaying charge current is also observed in open loop 
systems~\cite{butt2,akkermans,xia,mello,ore2,jaya4,xiong,jaya5,belu2},
where rings/cylinders are coupled to source and drain electrodes.
In $1985$, B\"{u}ttiker has introduced a conceptually simple 
and elegant approach~\cite{butt2} to investigate the effect of an 
electron reservoir on persistent current in a loop penetrated by a
magnetic flux. The reservoir is considered as a source of dissipation 
and the inelastic scattering processes take place only within this reservoir 
to which the loop is attached via a single current lead while the other 
scatterings occur in the lead are assumed to be elastic. So, there exists 
a complete spatial separation between the elastic and inelastic scatterings. 
The reservoir being a sink of electrons maintained the chemical potential of 
the loop to a fixed value. This modifies the statistical mechanical 
treatment of the system to a different ensemble, the grand canonical 
ensemble where close systems (in absence of lead and reservoir) belong 
to canonical ensemble average. This establishes a remarkable difference 
between the study of closed and open loop systems from the statistical point 
of view. Experimental verification of persistent current in open system 
was first done by Mailly {\em et al.} in 1993~\cite{mailly3}. They have
examined both closed and open systems and detected a periodic signal 
carrying the signature of persistent current which was in agreement with 
the theory. After those pioneering works last two decades have witnessed 
several approaches to reveal the properties of persistent currents in 
open systems. Unlike the behavior of the other physical quantity such 
as conductance, persistent current in open system is sensitive to the 
direction of the transport current and this property is helpful for 
recognizing this current from the other currents (noise) associated with 
experimental measurements. In open systems, persistent current can appear 
even in absence of external magnetic field~\cite{jaya6} as the lead-current, 
the current from the source to drain, plays the role of the driving force. 
Jayannavar {\em et al.}~\cite{jaya6,jaya7} have shown theoretically the 
flow of persistent current in an open metallic loop connecting to two 
electron reservoirs. They have taken the lengths of the two arms of the 
loop unequal which results in a circulating current in the loop. Instead 
of setting the arm-lengths unequal one can also introduce any local 
scatterer anywhere in the geometry to establish a circular flow within the 
loop geometry in non-equilibrium situation. On the other hand, in an
equilibrium condition i.e., when the chemical potentials of both leads 
are same, persistent current can still arise due to evanescent modes. In 
a work by Jayannavar {\it et al.} the phenomenon of persistent current due 
to two non-classical effects, AB effect and quantum mechanical tunneling 
has been studied in detail~\cite{jaya7,jaya8}. 

Although the studies involving persistent current in single-channel rings
and multi-channel rings~\cite{rg} or cylinders, both in the closed and 
open loop shapes, have already generated a wealth of literature there is 
still need to look deeper into the problem to address several significant 
issues those have not been well explored before as for examples the 
understanding of the distribution of persistent currents in different 
channels in presence of electron-electron interaction and disorder which 
inspect the net response of the full system and also the sensitivity of 
persistent current on disorder in partially disordered systems like 
shell-doped nanowires which can provide a unfamiliar electronic behavior 
rather than uniformly doped nanowires. In the present review we essentially 
concentrate on these issues.

In the first part we address magneto-transport properties in a zigzag
carbon nanotube, formed by rolling up a graphite ribbon in the cylindrical 
form~\cite{kit}, with its detailed energy band structure in presence 
electron-electron interaction within a Hartree-Fock (HF) mean field approach
using the second quantized formulation. Since the isolation of a single 
layer graphene by Novoselov {\it et al.}~\cite{geim1} intense and diverse 
research is going on to explore electronic transport in this system. Graphene, 
a single layer of carbon atoms tightly packed into a two-dimensional
honey-comb lattice, has drawn attention of scientists in various
disciplines due to its unconventional and fascinating electronic
properties arising particularly from the linear dispersion relation
around the Dirac points of the hexagonal Brillouin zone. These unique
properties can be understood in terms of the Dirac Hamiltonian~\cite{neto}
since it actually describes the physics of electrons near the Fermi level
of the undoped material. The carriers in graphene effectively behave as
massless relativistic particles within a low energy range close to Fermi
energy and these massless Dirac Fermions~\cite{geim2} evince various
phenomena in this energy range. The bipartite character of the wonderful
lattice structure of graphene strongly influences its intrinsic properties
and makes graphene a wonderful testbed for quantum field theory and
mathematical physics as well as condensed matter theory. In the last few
years extensive studies on persistent current in carbon nanotubes have been 
performed and many interesting physical phenomena have been
explored~\cite{sasaki,chen,szopa}. Persistent current in a carbon nanotube
is highly sensitive to its radius, chirality, deformation, etc. In a recent
experiment it has also been observed that the Fermi energy of a carbon 
nanotube can be regulated nicely by means of electron or hole doping, which 
can induce a dramatic change in persistent current~\cite{szopa}. It is well 
established that in a conventional multi-channel mesoscopic cylinder electron 
transport strongly depends on the correlation among different channels as 
well as the shape of Fermi surface. Therefore, we might expect some 
interesting features of persistent current in a carbon nanotube due to its 
unique electronic structure. Here we discuss the behavior of persistent 
current in different branches together with the total current of a zigzag 
carbon nanotube in presence of an AB flux. Most of the works available 
in literature~\cite{sasaki,chen,szopa,gefen,san3,san4,belu,ore1,peeters}
generally investigate magnetic response of the entire system, but a 
complete knowledge of magnetic response in individual branches provide
much better insight to the problem~\cite{we1}. The phase reversal of
persistent current in different branches as a function of electron-electron
correlation is found to exhibit interesting pattern. Our detailed numerical
analysis suggests a filling-dependent metal-insulator (MI) transition in 
a zigzag carbon nanotube.

In the second part, we concentrate on the behavior persistent current 
in an ordered-disordered separated nanotube considering a possible
experimental realization of shell-doped nanowires which may provide 
several unusual electronic behavior rather than uniformly doped 
nanowires and have potential applications in nanoscale electronic 
and optoelectronic devices. 

In the last part we investigate persistent current together with average 
density of states (ADOS) in an open loop geometry where an ordered binary 
alloy ring threaded by a magnetic flux is clamped between two ideal 
semi-infinite metallic electrodes, commonly known as source and drain 
electrodes, followed by the characteristic properties of an isolated 
ordered binary alloy ring. Inclusion of some foreign atoms in anyone of 
the two arms of the ring provides some interesting patterns in ADOS and 
from our numerical analysis we propose that under suitable choices of the 
parameter values the system can act as a $p$-type or an $n$-type 
semiconductor.

Throughout the review we choose $c=e=h=1$ for numerical calculations
and restrict ourselves at absolute zero temperature.

\section{Magneto-transport in a zigzag carbon nanotube}

This section is devoted to reveal the magnetic response of interacting 
electrons in a zigzag carbon nanotube enclosing a magnetic flux within a 
HF mean field approach. Following the description of energy 
\begin{figure}[ht]
{\centering \resizebox*{7cm}{5cm}{\includegraphics{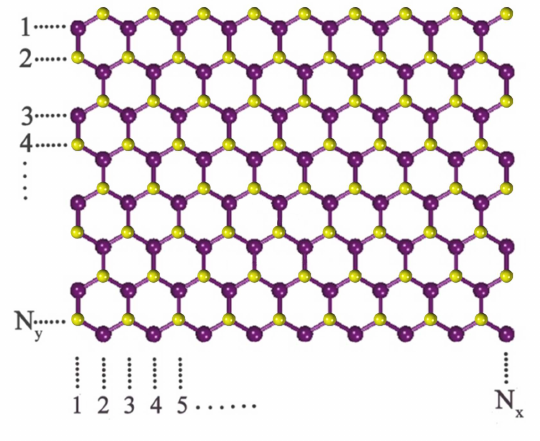}}\par}
\caption{(Color online). Schematic view of a zigzag graphite nano-ribbon 
with $N_x$ and $N_y$ number of atomic sites along the $x$ and $y$ 
directions, respectively.}
\label{zig}
\end{figure}
spectra for both non-interacting and interacting cases we investigate the 
energy-flux characteristics, persistent current in individual branches of 
the system and also the net current of the entire system.

\subsection{The model and the mean field scheme}

\begin{figure}[ht]
{\centering \resizebox*{3cm}{4.5cm}{\includegraphics{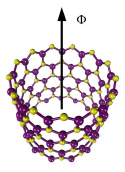}}\par}
\caption{(Color online). A graphite nanotube threaded by a magnetic flux 
$\Phi$.}
\label{tube}
\end{figure}

We begin by referring to Fig.~\ref{zig} where a graphite nano-ribbon of 
zigzag edges is shown. The filled magenta (large) and yellow (small) circles 
correspond to two different sub-lattices, namely, A and B, respectively. 
$N_x$ and $N_y$ are the number of atomic sites along the $x$ and $y$
directions, respectively. In order to elucidate magnetic response of a 
nanotube we roll up the graphite ribbon along $x$ direction using periodic 
boundary condition and allow to pass a magnetic flux $\phi$ (measured in 
unit of elementary flux quantum $\phi_0=c h/e$) along the axis of the tube 
as shown in Fig.~\ref{tube}. Our model quantum system is illustrated by the 
nearest-neighbor tight-binding (TB) framework which captures most of the 
essential properties of the tube nicely~\cite{sancho,lin,sorella,heyd}. 
To incorporate the effect of electron-electron interaction in the 
Hamiltonian we employ a HF mean field approximation. In Wannier basis, 
the Hamiltonian of an zigzag nanotube takes the form,
\begin{eqnarray}
\mbox{\boldmath $H$} & = & t \sum_{m,n,\sigma} 
\left(\mbox{\boldmath $a$}_{m,n,\sigma}^{\dag} 
\mbox{\boldmath $b$}_{m-1,n,\sigma} e^{-i \theta} \right. \nonumber \\ 
 & + & \left. \mbox{\boldmath $a$}_{m,n,\sigma}^{\dag} 
\mbox{\boldmath $b$}_{m+1,n,\sigma}\,e^{i \theta}
+ \mbox{\boldmath$a$}_{m,n,\sigma}^{\dag} 
\mbox{\boldmath$b$}_{m,n+1,\sigma}\right) + \mbox{h.c.} \nonumber \\
 & + & U \sum_{m,n} \left(\mbox{\boldmath$a$}_{m,n,\uparrow}^{\dag}
\mbox{\boldmath$a$}_{m,n\uparrow} \mbox{\boldmath$a$}_{m,n,\downarrow}^{\dag}
\mbox{\boldmath$a$}_{m,n,\downarrow} \right. \nonumber \\
 & + & \left. \mbox{\boldmath$b$}_{m+1,n,\uparrow}^{\dag}
\mbox{\boldmath$b$}_{m+1,n,\uparrow} 
\mbox{\boldmath$b$}_{m+1,n,\downarrow}^{\dag} 
\mbox{\boldmath$b$}_{m+1,n,\downarrow} \right)
\label{ham}
\end{eqnarray}
where, $\mbox{\boldmath$a$}_{m,n,\sigma}^{\dag}$ 
($\mbox{\boldmath$b$}_{m,n,sigma}^{\dag}$) is the creation operator for an 
\begin{figure}[ht]
{\centering \resizebox*{3cm}{3cm}{\includegraphics{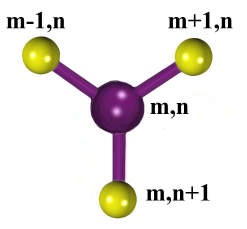}}\par}
\caption{(Color online). Schematic view of different atomic sites
with their co-ordinates.}
\label{index}
\end{figure}
up spin or down spin electron of spin associated with A (B) type of atoms 
at the position ($m$,$n$) and the corresponding annihilation operator is 
denoted by $a_{m,n}$ ($b_{m,n}$). $t$ is the nearest-neighbor hopping 
integral and $U$ is the strength of on-site Hubbard interaction. The 
co-ordinates of the lattice sites are denoted by two integers, $m$ and $n$. 
The site indexing is schematically shown in Fig.~\ref{index} for
better viewing. The factor $\theta$ ($=2 \pi \phi/N_x$), the so-called 
Peierl's~\cite{peierls} phase factor, is introduced into the above 
Hamiltonian to incorporate the effect of magnetic flux applied along 
the axis of the tube.
\vskip 0.2cm
\noindent
{\em \underline {Decoupling of interacting Hamiltonian}}:
Using the generalized HF approach, the so-called mean-field approximation, 
we decouple the TB Hamiltonian into three different parts two of which 
correspond to two different values of $\sigma$, $\uparrow$ and $\downarrow$, 
with modified site-energies. The decoupled Hamiltonian is expressed as,
\begin{equation}
\mbox {\boldmath{$H$}}_{\mbox{\tiny MF}}=\mbox {\boldmath{$H$}}_{\uparrow}
+\mbox {\boldmath{$H$}}_{\downarrow}+\mbox {\boldmath{$H$}}_0
\end{equation}
where,
\begin{eqnarray}
\mbox{\boldmath$H$}_{\uparrow} & = & U \sum_{m,n} \left(\langle 
\mbox{\boldmath$n$}^a_{m,n,\downarrow}\rangle
\mbox{\boldmath$n$}^a_{m,n,\uparrow} + 
\langle \mbox{\boldmath$n$}^b_{m+1,n,\downarrow}\rangle 
\mbox{\boldmath$n$}^b_{m+1,n,\uparrow} \right) \nonumber \\
 & + & t \sum_{m,n} \left(\mbox{\boldmath$a$}_{m,n,\uparrow}^{\dag}
\mbox{\boldmath$b$}_{m-1,n,\uparrow} e^{-i \theta} 
+ \mbox{\boldmath$a$}_{m,n,\uparrow}^{\dag} 
\mbox{\boldmath$b$}_{m+1,n,\uparrow} e^{i \theta} 
\right.  \nonumber \\
 & + & \left. \mbox{\boldmath$a$}_{m,n,\uparrow}^{\dag} 
\mbox{\boldmath$b$}_{m,n+1,\uparrow} + \mbox{h.c.} \right), \nonumber \\
\mbox{\boldmath$H$}_{\downarrow}&=& U \sum_{m,n} \left(\langle 
\mbox{\boldmath$n$}^a_{m,n,\uparrow}\rangle 
\mbox{\boldmath$n$}^a_{m,n,\downarrow} 
+ \langle \mbox{\boldmath$n$}^b_{m+1,n,\uparrow}\rangle 
\mbox{\boldmath$n$}^b_{m+1,n,\downarrow}\right) \nonumber \\
 & + & t \sum_{m,n} \left( \mbox{\boldmath$a$}_{m,n,\downarrow}^{\dag}
\mbox{\boldmath$b$}_{m-1,n,\downarrow} e^{-i \theta} 
+ \mbox{\boldmath$a$}_{m,n,\downarrow}^{\dag} 
\mbox{\boldmath$b$}_{m+1,n,\downarrow} e^{i \theta} \right. \nonumber \\
 & + & \left. \mbox{\boldmath$a$}_{m,n,\downarrow}^{\dag} 
\mbox{\boldmath$b$}_{m,n+1,\downarrow}+ \mbox{h.c.} \right), \nonumber \\ 
\mbox{\boldmath$H$}_0 &=&-U \sum_{m,n}\left(\langle 
\mbox{\boldmath$n$}^a_{m,n,\uparrow}\rangle 
\langle \mbox{\boldmath$n$}^a_{m,n,\downarrow}\rangle \right. \nonumber \\
 & + & \left. \langle \mbox{\boldmath$n$}^b_{m+1,n,\uparrow}\rangle 
\langle \mbox{\boldmath$n$}^b_{m+1,n,\downarrow}\rangle \right).
\end{eqnarray}
where,  $\mbox{\boldmath$H$}_{\uparrow}$ and 
$\mbox{\boldmath$H$}_{\downarrow}$ represent the up-spin and down-spin 
Hamiltonians, respectively. $\mbox{\boldmath$H$}_0$ is a constant term 
which gives the energy shift. Here, $\mbox{\boldmath$n$}^a_{m,n,\sigma}$ 
and $\mbox{\boldmath$n$}^b_{m,n,\sigma}$ are the number operators 
associated with the A and B sites, respectively. 
\vskip 0.2cm
\noindent
{\em \underline{Self-consistent procedure}}:
In order to get the energy eigenvalues of the interacting Hamiltonian we go
through a self-consistent procedure considering initial guess values of
$\langle n^a_{m,n,\sigma}\rangle$ and $\langle n^b_{m,n,\sigma}\rangle$.
With these initial values, the up and down spin Hamiltonians are
diagonalized numerically and a new set of values of
$\langle n^a_{m,n,\sigma}\rangle$ and $\langle n^b_{m,n,\sigma}\rangle$
are calculated. These steps are repeated until a self-consistent solution
is achieved.
\vskip 0.2cm
\noindent
{\em \underline{Finding the ground state energy}}:
After getting the self-consistent solution we determine the ground state
energy ($E_0$) at absolute zero temperature ($T=0\,K$) for a particular
filling by taking the sum of individual states upto the Fermi level
($E_F$) for both up and down spin electrons. The expression for ground
state energy reads,
\begin{equation}
E_0=\sum_{i}E_{i,\uparrow}+\sum_{i}E_{i,\downarrow}+H_0
\end{equation}
where, $i$ runs over the states up to the Fermi level. $E_{i,\uparrow}$'s
and $E_{i,\downarrow}$'s are the single particle energy eigenvalues obtained
by diagonalizing the up and down spin Hamiltonians 
$\mbox{\boldmath$H$}_{\uparrow}$ and $\mbox{\boldmath$H$}_{\downarrow}$, 
respectively.

\subsection{Energy band structure}

To make this present communication a self contained study let us first
start with the energy band structure of a finite width zigzag nano-ribbon.
\vskip 0.2cm
\noindent
{\em \underline{Non-interacting case}}:
To establish the energy dispersion relation of a zigzag nano-ribbon
we find an effective difference equation analogous to the case of an
infinite one-dimensional chain.
\begin{figure}[ht]
{\centering \resizebox*{3.5cm}{5cm}{\includegraphics{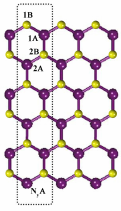}}\par}
\caption{(Color online). Unit cell configuration of a zigzag nano-ribbon.}
\label{unit}
\end{figure}
This can be done by proper choice of a unit cell from the nano-ribbon. 
The schematic view of the unit cell configuration with $N_y$ pairs of 
B-A atoms in a zigzag nano-ribbon is shown in Fig.~\ref{unit}. With this 
arrangement, we express the effective difference equation of the 
nano-ribbon in the form,
\begin{equation}
(E \,\mathcal{I} - \mathcal{E}_{\sigma})\psi_{j,\sigma} = \mathcal{T} 
\psi_{j+1,\sigma} + \mathcal{T}^{\dag} \psi_{j-1,\sigma} 
\label{diff}
\end{equation}
where, 
\begin{eqnarray}
\psi_{j,\sigma}= \left(\begin{array}{c}
\psi_{j1B,\sigma} \\
\psi_{j1A,\sigma} \\
\psi_{j2B,\sigma} \\
.\\
.\\
\psi_{jN_yA,\sigma}\end{array}\right).
\end{eqnarray}
$\mathcal{E}$ and $\mathcal{T}$ are the site-energy and nearest-neighbor 
hopping matrices of the unit cell, respectively. $\mathcal{I}$ is a
($2N_y\times2N_y$) identity matrix. According to our convention, the
translational invariance of the nano-ribbon exists along the $x$-direction
and we can write $\psi_{j,\sigma}$ in terms of the Bloch waves and then 
Eq.~\ref{diff} takes the form,
\begin{equation}
(E \mathcal{I}-\mathcal{E}_{\sigma})=\mathcal{T} e^{ik_x \Lambda}+
\mathcal{T}^{\dag} e^{-ik_x \Lambda}
\label{bloch}
\end{equation}
where, $\Lambda=\sqrt{3} a$ is the horizontal separation between two
filled magenta or yellow circles situated at two successive unit cells.
$a$ is the length of each side of a hexagonal benzene like ring. Finally,
we solve Eq.~\ref{bloch} to get the desired energy dispersion relation 
($E$ vs. $k_x$) of the ribbon.

\begin{figure}[ht]
{\centering \resizebox*{7.5cm}{4cm}{\includegraphics{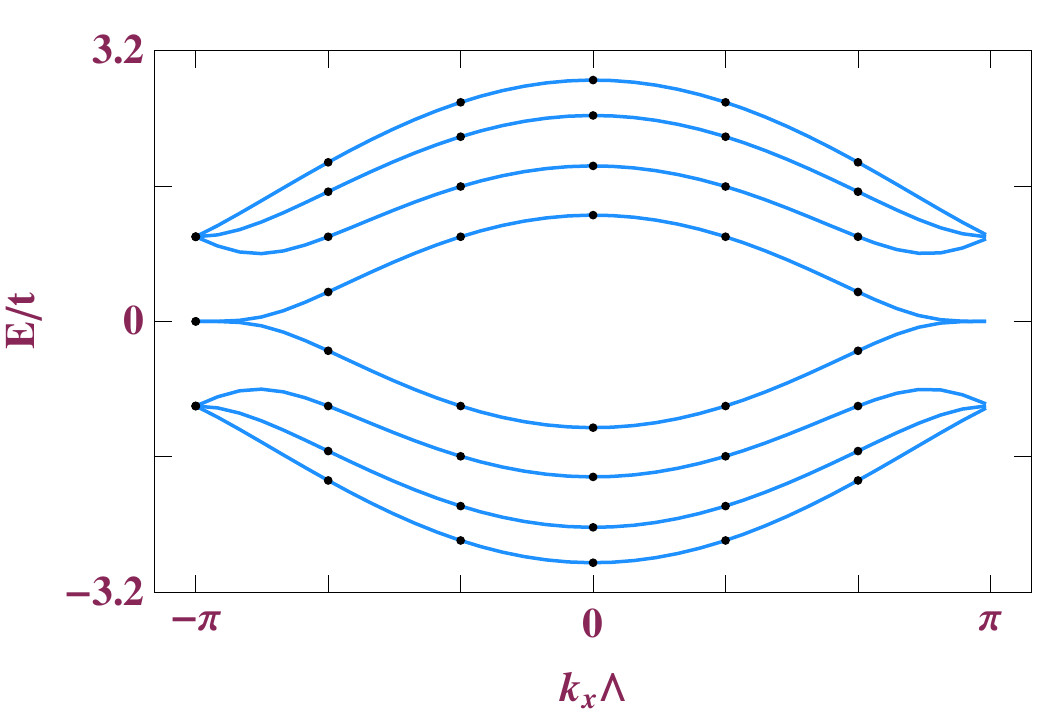}}\par}
\caption{(Color online). Energy levels (blue curve) as function of $k_x$ 
for a finite width zigzag nano-ribbon considering $N_y=4$. The discrete 
eigenvalues (filled black circles) of a nanotube with $N_x=12$ and $N_y=4$, 
in absence of AB flux $\phi$, are superimposed. Here we set $U=0$.}
\label{band}
\end{figure}

As illustrative example in Fig.~\ref{band}, we display the variation of 
energy levels (blue curves) as a function of wave vector $k_x$ for a finite
width zigzag nano-ribbon considering $N_y=4$. At $E=0$, nearly flat bands 
appear in the spectrum. The electronic states corresponding to those almost 
flat bands are strongly localized near the zigzag edges of the tube. The 
existence of these edge states have also been reported earlier by some 
other groups~\cite{waka1,brey,neto}. 

With this energy band structure of a finite width nano-ribbon we now pay 
attention on the variation of energy levels of a nanotube. In order to get 
the nanotube from the ribbon we apply periodic boundary condition along the 
$x$-direction~\cite{waka2} which results quantized values of $k_x$, and 
the quantized wave numbers are expressed from the relation $k_x=4\pi n_x/
N_x \Lambda$, where $n_x$ is an integer lies within the range:
$-N_x/4\leq n_x< N_x/4$. Plugging the quantized values of $k_x$ in
Eq.~\ref{bloch} we can easily determine the eigenvalues of a finite sized
nanotube. In Fig.~\ref{band} we represent the discrete energy eigenvalues 
(filled black circles) for a zigzag nanotube considering $N_x=12$ and 
$N_y=4$. It is to be noted that the results displayed in Fig.~\ref{band} 
correspond to the case when AB flux $\phi$ is set equal to zero. With 
these parameter values of the nanotube $k_x$ gets six quantized values 
($-\pi/\Lambda$, $-2\pi/3\Lambda$, $-\pi/3\Lambda$, $0$, $\pi/3\Lambda$ 
and $2\pi/3\Lambda$), and therefore, total $48$ energy values are obtained
since $N_y$ is fixed at $4$.

\vskip 0.2cm
\noindent
{\em \underline{Interacting case}}: In the presence of e-e interaction
energy levels get modified significantly depending on the filling of
\begin{figure}[ht]
{\centering \resizebox*{7.5cm}{7cm}{\includegraphics{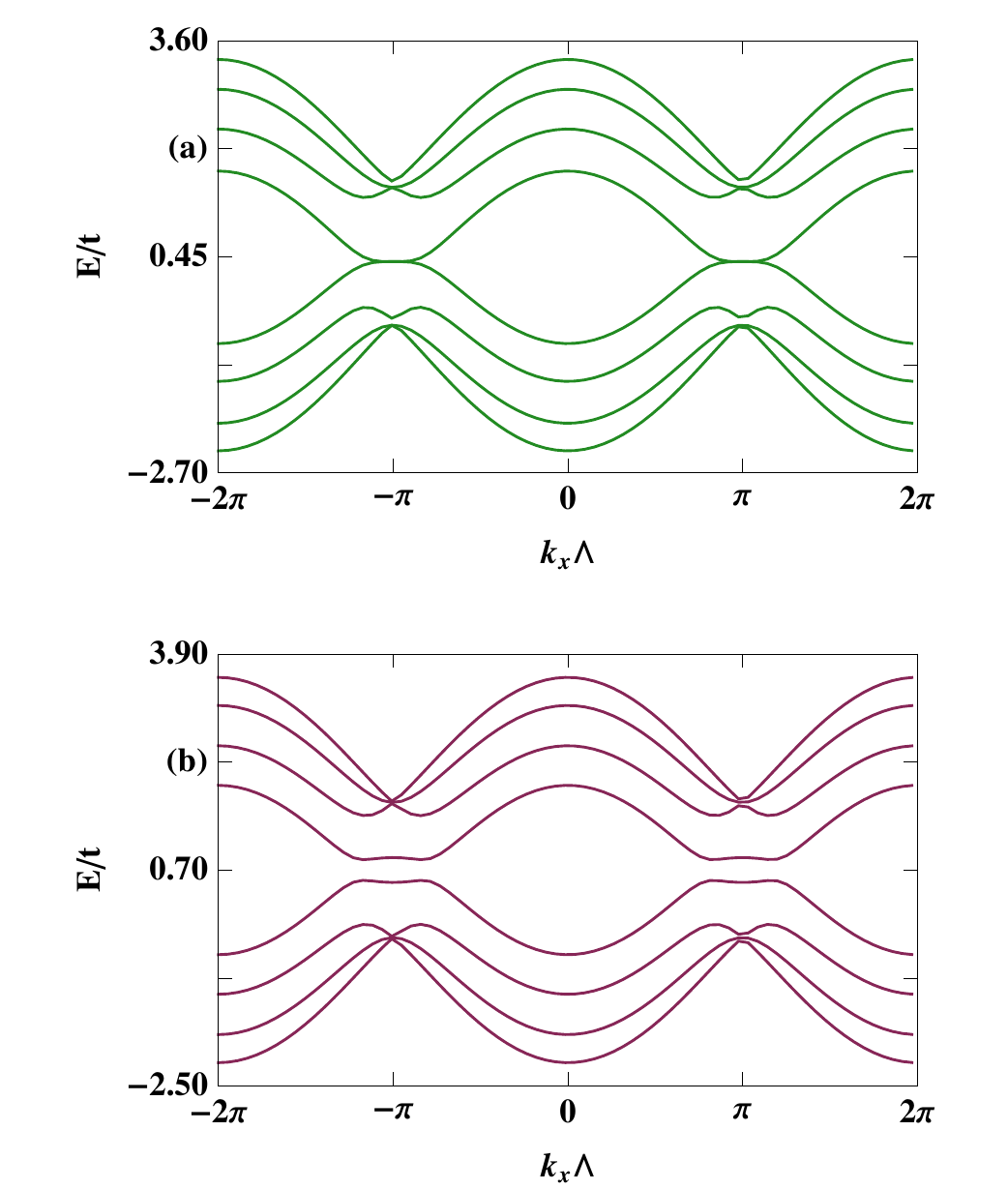}}\par}
\caption{(Color online). Energy levels as function of $k_x$ for a finite
width zigzag nano-ribbon considering $N_y=4$ and $U=1.4$, where (a) and
(b) correspond to the one-third- and half-filled cases, respectively.}
\label{uband}
\end{figure}
electrons. The results calculated for a particular value of $U$ are
presented in Fig.~\ref{uband} where we set $N_y=4$. In the half-filled
band case, a gap opens up at the Fermi energy~\cite{rossier} which is
consistent with the DFT calculations~\cite{son} and the gap increases
with the value of $U$. A careful investigation also predicts that the
full energy band gets shifted by the factor $U/2$.

\subsection{Energy-flux characteristics}

\begin{figure}[ht]
{\centering \resizebox*{7.5cm}{7cm}{\includegraphics{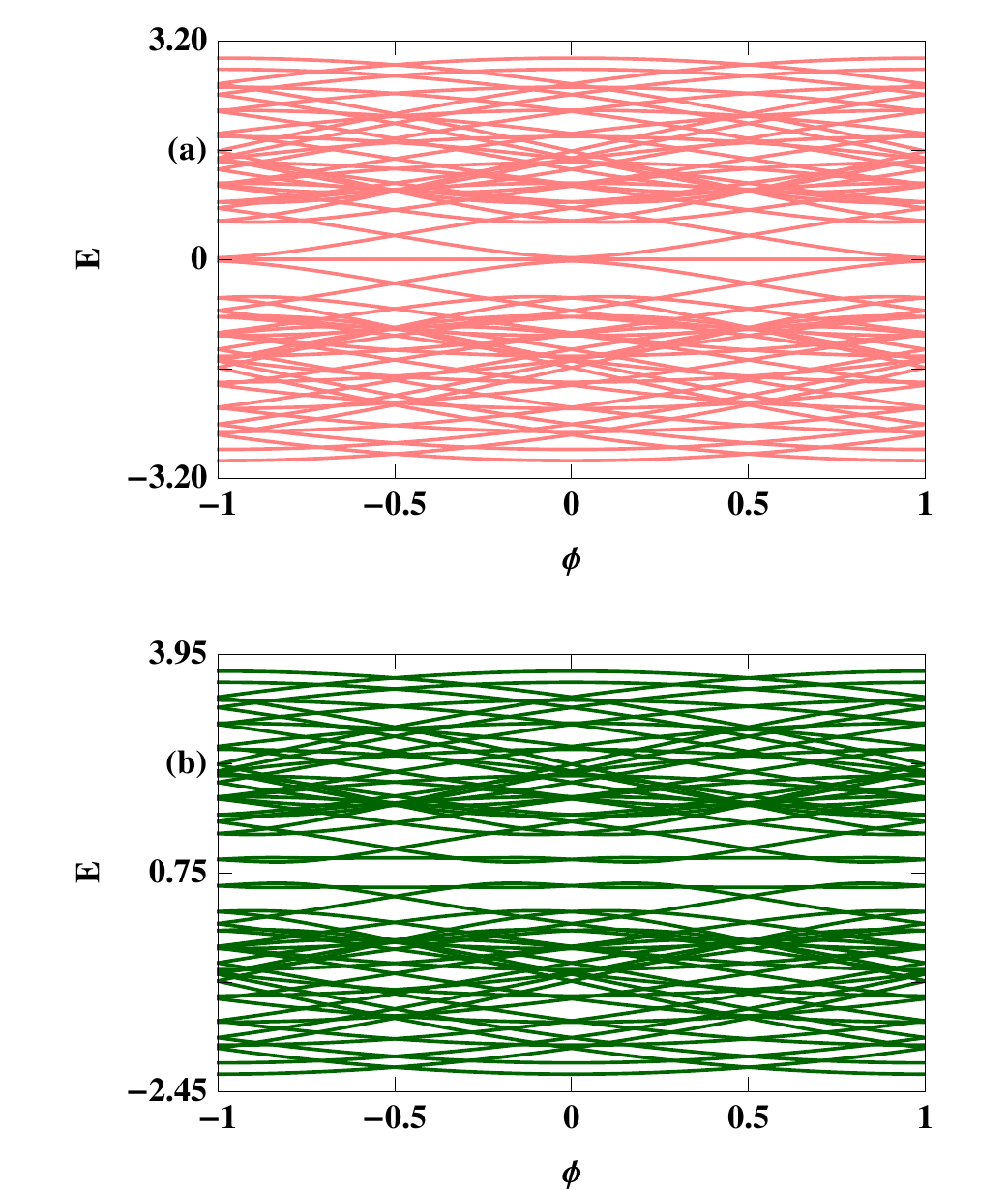}}\par}
\caption{(Color online). Energy-flux characteristics of a half-filled zigzag 
nanotube with $N_x=10$ and $N_y=7$. (a) $U=0$ and (b) $U=1.5$.}
\label{energy}
\end{figure}

In this sub-section we examine the energy-flux characteristics of a zigzag
nanotube. The results for a zigzag nanotube considering $N_x=10$ and $N_y=7$
are displayed in Fig.~\ref{energy}, where (a) and (b) represent $U=0$ and 
$U=1.5$, respectively.
In the absence of Hubbard interaction ($U=0$), energy levels are obtained 
simply by diagonalizing the non-interacting Hamiltonian and the nature of 
the energy spectrum becomes independent of the total number of electrons 
$N_e$ in the system. While, for the interacting case ($U \ne 0$) we employ a 
mean-field scheme where the interacting Hamiltonian (Eq.~\ref{ham}) are 
decoupled (for a particular filling) into two non-interacting Hamiltonians 
corresponding to up and down spin electrons and then diagonalize both
$\mbox {\boldmath{$H$}}_{\uparrow}$ and $\mbox {\boldmath{$H$}}_{\downarrow}$ 
to get the energy eigenvalues of the system. Since in our case we set 
$N_x=10$ and $N_y=7$, we get total $70$ independent energy levels and due 
to their overlaps individual energy levels are not clearly distinguished 
from the spectra given in Fig.~\ref{energy}. For identical filling factor 
of up and down spin electrons the energy levels are exactly similar both 
for $\mbox {\boldmath{$H$}}_{\uparrow}$ and 
$\mbox {\boldmath{$H$}}_{\downarrow}$ (see Fig.~\ref{energy}(b)), and 
therefore, one energy spectrum cannot be separated from the other. At $E=0$, 
the energy levels become almost flat 
for a wide range of $\phi$, and, near $\phi=\pm \phi_0/2$ they vary slowly 
with $\phi$ as shown in Fig.~\ref{energy}(a). In Fig.~\ref{energy}(b) the 
variation of energy levels with $\phi$ for a zigzag nanotube with the same 
parameter values sated above are plotted considering Hubbard interaction. 
Here we choose $U=1.5$. Both for the up and down spin Hamiltonians the 
eigenvalues are exactly identical and they overlap with 
\begin{figure}[ht]
{\centering \resizebox*{6.7cm}{4cm}{\includegraphics{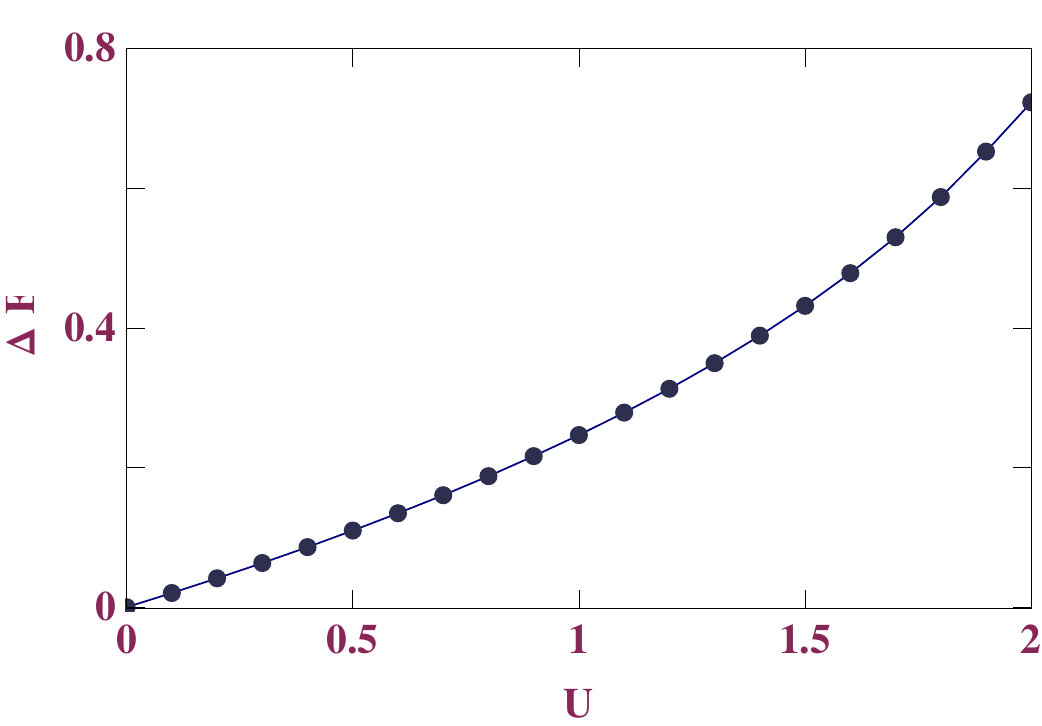}}\par}
\caption{(Color online). Energy gap ($\Delta E$) as a function of on-site 
Hubbard interaction strength $U$ for a zigzag nanotube with $N_x=10$ 
and $N_y=7$ in the half-filled band case when $\phi$ is set at $\phi_0/2$.}
\label{gap}
\end{figure}
each other. In absence of interaction there is no gap in the center 
(around $E=0$) of the band spectrum (Fig.~\ref{energy}(a)). As soon as 
the interaction is taken into consideration a finite gap opens up in the
band center (Fig.~\ref{energy}(b)) and it increases with the strength of 
interaction ($U$) as shown in Fig~\ref{gap}. This energy gap is consistent 
with the energy gap obtained in the $E$-$k_x$ diagram (Fig.~\ref{uband}(b)). 
All these energy levels vary
periodically with $\phi$ providing $\phi_0$ ($=1$ in our chosen unit) 
flux-quantum periodicity. At half-integer or integer multiples of $\phi_0$, 
energy levels have either a maximum or a minimum (see Fig.~\ref{energy}). 
Both the energy spectra take a complicated look as there are many crossings 
among the energy levels particularly in the regions away from $E=0$ 
(Fig.~\ref{energy}). 

\begin{figure}[ht]
{\centering \resizebox*{8.2cm}{6cm}{\includegraphics{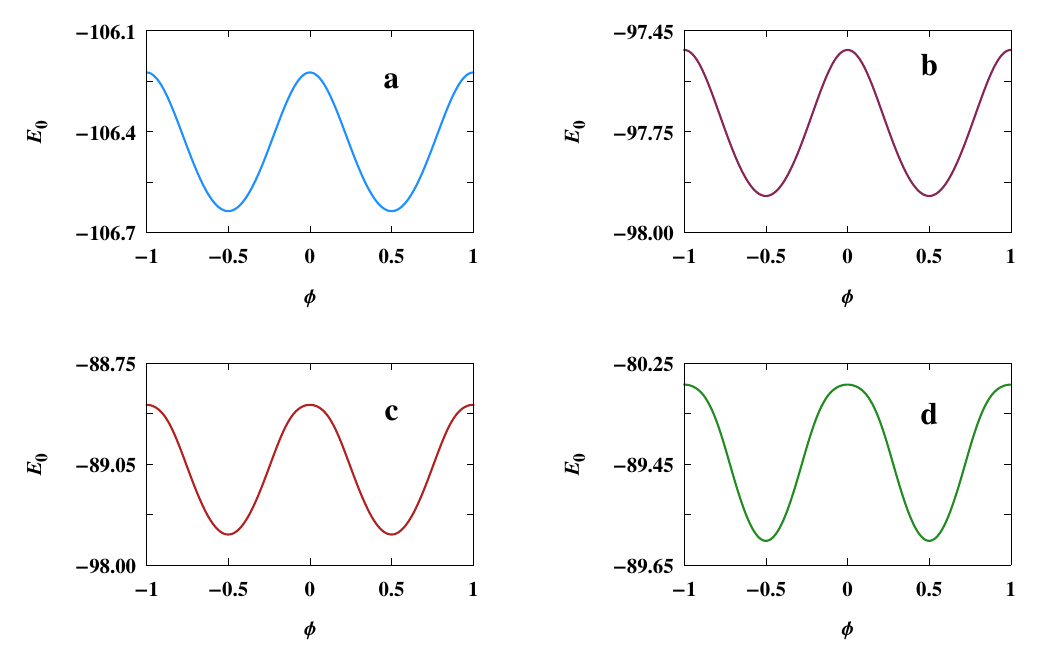}}\par}
\caption{(Color online). Ground state energy level as a function of $\phi$
for a zigzag nanotube in the half-filled band case considering $N_x=20$ and
$N_y=8$. (a), (b), (c) and (d) correspond to $U=0$, $0.5$, $1$ and $1.5$,
respectively.}
\label{ground}
\end{figure}

In Fig.~\ref{ground} we present the variation of ground state energy of a 
carbon nanotube with zigzag edges as a function of magnetic flux $\phi$ for 
the half-filled band case. Here (a)-(d) represent the four different cases 
corresponding to four different values of electronic correlation strength 
$U=0$, $0.5$, $1$ and $1.5$, respectively. The energy levels evince one 
flux-quantum periodicity, as expected, and their energies get increased 
with $U$.  

\subsection{Evaluation of persistent current in the second quantized form}

Following the second quantized formulation~\cite{san5,san6}, we estimate 
persistent current in individual zigzag paths of a nanotube, threaded by 
an AB flux $\phi$. This is an elegant and nice way of studying the 
response in separate branches of any quantum network.

At first, we express the basic equation of current operator 
\mbox{\boldmath$I_{\sigma}$} corresponding to spin $\sigma$ in terms of 
the velocity operator \mbox{\boldmath$v_{\sigma}$}
(=\mbox{\boldmath ${\dot{x}_{\sigma}}$}) as,
\begin{equation}
\mbox{\boldmath $I_{\sigma}$}=-\frac{1}{N_x} e 
\mbox{\boldmath ${\dot{x}_{\sigma}}$}
\label{velo}
\end{equation}
where, \mbox{\boldmath ${x_{\sigma}}$} is the displacement operator. The 
velocity operator is obtained from the relation,
\begin{eqnarray}
\mbox{\boldmath $v_{\sigma}$} &=& \frac{1}{i\hbar}
\left[\mbox{\boldmath ${x_{\sigma}}$},\mbox{\boldmath ${H_{\sigma}}$}\right].
\label{equ33}
\end{eqnarray} 
We use this expression to find the velocity operator of an electron
with spin $\sigma$ in a zigzag  channel $n$ (say) in the form,
\begin{eqnarray}
\mbox{\boldmath$v_{n,\sigma}$}& = &\frac{t}{i \hbar} \sum_m 
\left(\mbox{\boldmath$b$}^{\dag}_{m+1,n,\sigma} 
\mbox{\boldmath$a$}_{m,n,\sigma}e^{-i \theta} \right. \nonumber \\
 & & - \left. \mbox{\boldmath$a$}^{\dag}_{m,n,\sigma} 
\mbox{\boldmath$b$}_{m+1,n,\sigma}e^{i \theta} 
-\mbox{\boldmath$b$}^{\dag}_{m-1,n,\sigma}
\mbox{\boldmath$a$}_{m,n,\sigma}e^{i \theta} \right. \nonumber \\
 & & + \left. \mbox{\boldmath$a$}^{\dag}_{m,n,\sigma} 
\mbox{\boldmath$b$}_{m-1,n,\sigma}e^{-i \theta} \right).
\label{velop}
\end{eqnarray}
For a particular eigenstate $|\psi_{p,\sigma}\rangle$ persistent current in 
$n$-th channel becomes, 
\begin{equation}
I_{n,\sigma}^p=-\frac{e}{N_x}\langle \psi_{p,\sigma}|
\mbox{\boldmath$v_{n,\sigma}$}|\psi_{p,\sigma}\rangle
\label{currequ}
\end{equation}
where, the eigenstate $|\psi_{p,\sigma}\rangle$ is written as,
\begin{eqnarray}
|\psi_{p,\sigma} \rangle &=&\sum_{m,n} \left( \alpha_{m,n,\sigma}^p |
m,n,\sigma\rangle +\beta_{m-1,n,\sigma}^p |m-1,n ,\sigma\rangle 
\right. \nonumber \\
& & \left. + \beta_{m+1,n,\sigma}^p |m+1,n,\sigma \rangle 
+ \beta_{m,n+1,\sigma}^p |m,n+1,\sigma \rangle \right), \nonumber \\
\label{shi}
\end{eqnarray}
where, $|m,n,\sigma\rangle$'s are the Wannier states and 
$\alpha_{m,n,\sigma}^p$ and $\beta_{m,n,\sigma}^p$'s are the corresponding 
coefficients. Simplifying Eq.~\ref{currequ}, we get the final relation of 
persistent charge current for $n$-th zigzag channel as,
\begin{eqnarray}
I_{n,\sigma}^p & = & \frac{iet}{\hbar N_x} \sum_m
\left(\beta^{p~*}_{m+1,n,\sigma} \alpha^p_{m,n,\sigma}e^{-i \theta} 
\right. \nonumber \\
 & & \left. -\alpha^{p~*}_{m,n,\sigma}\beta^p_{m+1,n,\sigma}e^{i \theta}
- \beta^{p~*}_{m-1,n,\sigma} \alpha^p_{m,n,\sigma}e^{i \theta} 
\right. \nonumber \\
 & & \left. + \alpha^{p~*}_{m,n,\sigma}\beta^p_{m-1,n,\sigma}e^{-i \theta} 
\right).
\label{vel}
\end{eqnarray} 
With the above prescription we can also evaluate persistent current in
individual armchair paths (along $y$ direction) of the nanotube. Finally,
it takes the form,
\begin{widetext}
\begin{eqnarray}
I_{m-1,m,\sigma}^p & = &\frac{i t}{2 \hbar N_y}\left[\sum_{n=1,3,\hdots}^{N_y} 
\left(\alpha^{p~*}_{m,n,\sigma} \beta^p_{m-1,n,\sigma}e^{-i \theta} 
- \beta^{p~*}_{m-1,n},\sigma \alpha^p_{m,n,\sigma}e^{i \theta}
\right) 
+ \sum_{n=2,4,\hdots}^{N_y}\left(\beta^{p~*}_{m,n,\sigma} 
\alpha^p_{m-1,n,\sigma} e^{-i \theta} - \alpha^{p~*}_{m-1,n,\sigma} \right. 
\right. \nonumber \\
& \times &  \beta^p_{m,n,\sigma} e^{i \theta} 
+ \left. \beta^{p~*}_{m-1,n,\sigma} \alpha^p_{m-1,n-1,\sigma}
-\alpha^{p~*}_{m-1,n-1,\sigma} \beta^p_{m-1,n,\sigma} \right)
+ \left. \sum_{n=2,4,\hdots}^{N_y-1} \left( \beta^{p~*}_{m,n+1,\sigma} 
\alpha^p_{m,n,\sigma}-\alpha^{p~*}_{m,n,\sigma} \beta^p_{m,n+1,\sigma} \right)
\right] \nonumber \\
\label{vely}
\end{eqnarray}
\end{widetext}
where, an armchair channel ($m-1,m$) is constructed by ($m-1$)-th and 
$m$-th lines according to our indexing. It is noteworthy to mention that 
all the calculations are done at absolute zero temperature ($T=0\,K$). 
Now, the net persistent current driven by electrons of spin $\sigma$ in a 
particular channel $n$ for a nanotube described with Fermi energy $E_F$ 
can be determined by taking the sum of individual contributions from the 
lowest energy eigenstates upto the Fermi level. Therefore, we have,
\begin{equation}
I_{n,\sigma}=\sum_{p} I_{n,\sigma}^p.
\label{pcc}
\end{equation}
Taking the contributions from all possible channels $n$, both up and down 
spin ($\sigma$) electrons, the total persistent current in the nanotube 
can be expressed as,
\begin{equation}
I_T=\sum_{n,\sigma} I_{n,\sigma}.
\end{equation}
To judge the accuracy of the persistent current calculated from the present 
scheme we determine persistent current in some other ways as 
available in literature. Probably the simplest way of determining persistent 
current is the case where first order derivative of ground state energy with 
respect to AB flux $\phi$ is taken into account~\cite{san1,san2}. Therefore, 
we can write,
\begin{equation}
I_T=-c\frac{\partial E_0(\phi)}{\partial \phi}
\label{deri}
\end{equation}
where, $E_0(\phi)$ is the total ground state energy for a particular 
electron filling. But, in our present method, the so-called second 
quantized approach, there are some advantages compared to other available 
procedures. With the second quantized formulation we can easily determine 
current in any branch of a complicated network and the evaluation of 
individual responses in separate branches helps us to elucidate the actual 
mechanism of electron transport in a more transparent way.

\subsection{Current-flux characteristics}

Now in this subsection we focus our attention on the behavior of persistent 
current in a zigzag nanotube. Here, also we adopt the unit where $c=h=e=1$, 
fix $t=-1$ and measure all the physical quantities in unit of $t$. 

\begin{figure*}[ht]
{\centering \resizebox*{17cm}{12cm}
{\includegraphics{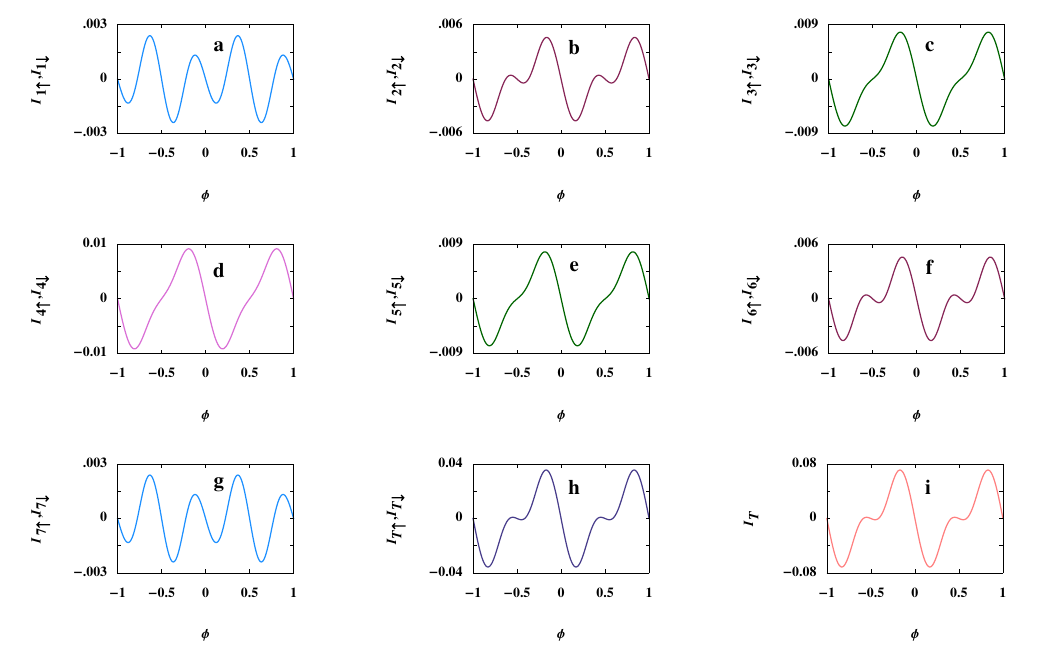}}\par}
\caption{(Color online). Persistent current in individual zigzag paths as a
function of $\phi$ for a half-filled zigzag nanotube ($N_x=20$ and $N_y=7$)
with $U=1.2$, where, (a)-(g) correspond to $1$st-$7$th zigzag channels of
the tube, respectively. The net current corresponding to both up and down
spin electrons are displayed in (h) while in (i) total persistent current is
shown.}
\label{udistribution}
\end{figure*}

\begin{figure}[ht]
{\centering \resizebox*{6.5cm}{4cm}{\includegraphics{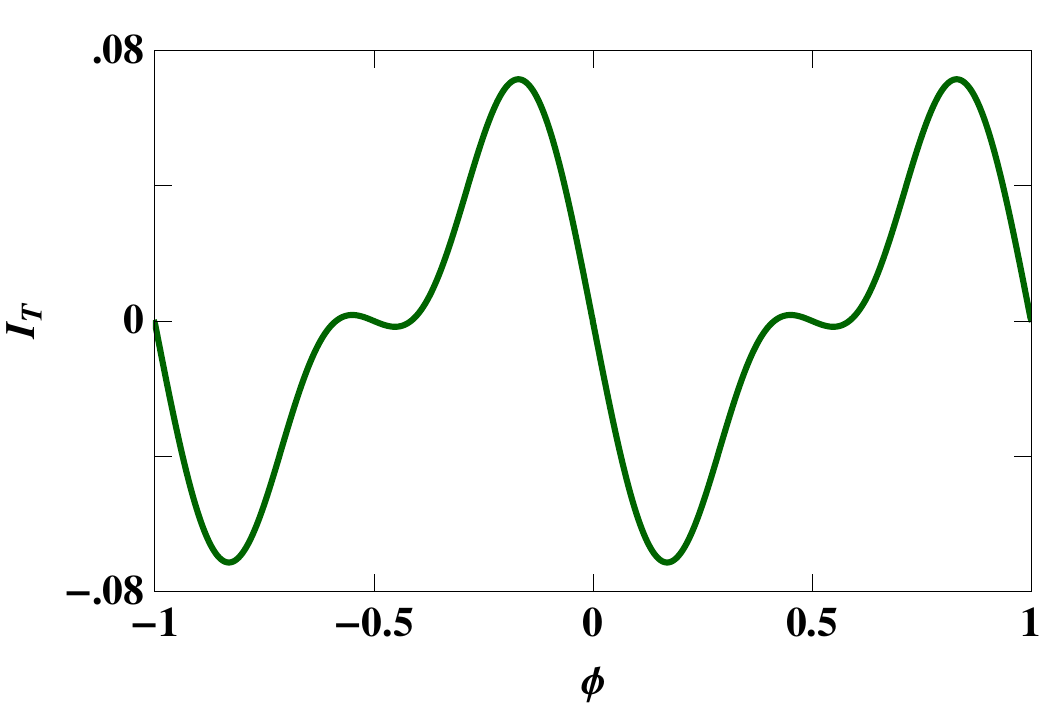}}\par}
\caption{(Color online). Total persistent current obtained in a traditional 
derivative approach (Eq.~\ref{deri}) as a function of $\phi$ for the same 
parameter values mentioned in Fig.~\ref{udistribution}.}
\label{total}
\end{figure}

\begin{figure}[ht]
{\centering \resizebox*{6.5cm}{4cm}{\includegraphics{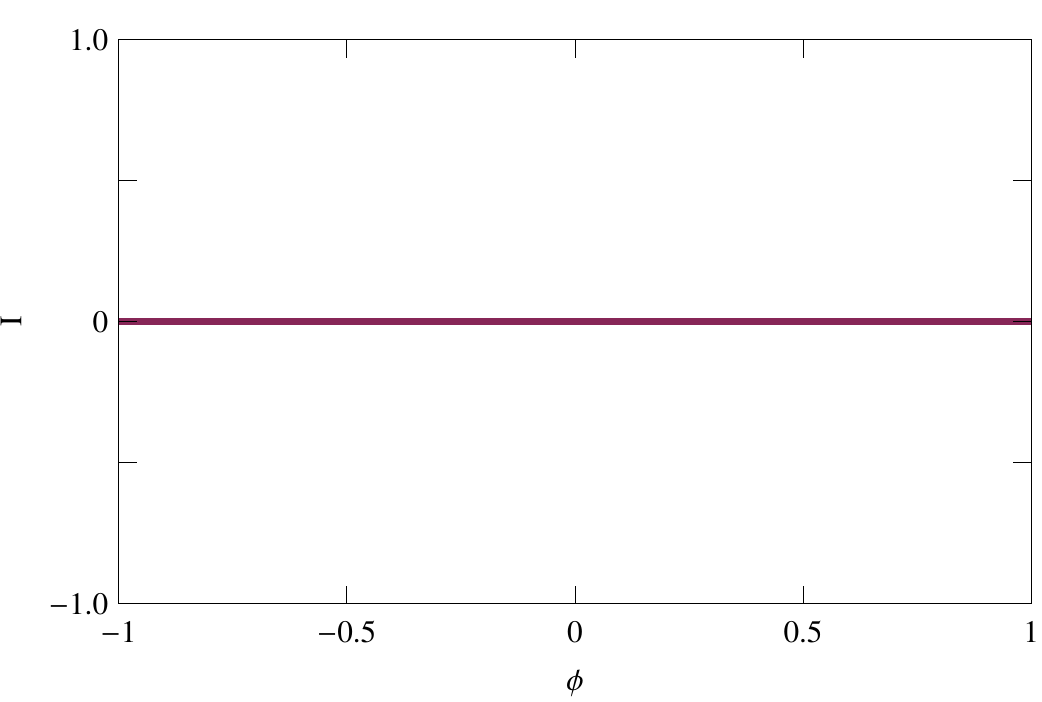}}\par}
\caption{(Color online). Persistent current in an armchair path as a 
function of $\phi$ for a zigzag nanotube with the same parameter values as 
mentioned in Fig.~\ref{udistribution}.}
\label{yvelocity}
\end{figure}

To illustrate the behavior of persistent current in separate branches of a 
zigzag carbon nanotube we show in Fig.~\ref{udistribution} the variation of 
persistent current in individual zigzag paths as a function of flux $\phi$ 
for the half-filled case considering $N_x=20$ and  $N_y=7$, where (a)-(g) 
correspond to $1$st-$7$th zigzag channels of the tube, respectively. The 
Hubbard correlation strength $U$ is fixed at $1.2$. Currents carried by 
up and down spin electrons are displayed in each of these figures 
simultaneously and they are exactly superposed on each other as the 
magnitudes and behaviors are exactly similar to each other. All these 
current profiles exhibit $\phi_0$ flux-quantum periodicity. There are 
no similarity in magnitudes of persistent currents corresponding to 
separate channels but also another similarity carried by them. 
Interestingly we observe that $I_{1\uparrow}$ is exactly identical to 
$I_{7\uparrow}$, and, similarly for the ($I_{2\uparrow}$, $I_{6\uparrow}$) 
and ($I_{3\uparrow}$, $I_{5\uparrow}$) pairs. $I_{4,\uparrow}$, the current 
in the middle channel, becomes the isolated one since we have chosen $N_y=7$. 
This is true for any zigzag nanotube with odd $N_y$. For a tube with even 
$N_y$, currents are pairwise identical. Summing up the individual currents 
in seven zigzag channels we get separately the net persistent current 
carried by up and down spin electrons in the nanotube which is presented in 
Fig.~\ref{udistribution}(h) and the total current is obtained by taking both
contributions $I_{T\uparrow}$ and $I_{T\downarrow}$ and it is depicted in 
Fig.~\ref{udistribution}(i). Now the total current derived from the 
conventional method where first order derivative of the ground state energy 
is taken into account, is displayed in Fig.~\ref{total}. These two net 
currents calculated from the two different schemes are exactly identical 
to each other. It emphasizes that the net contribution of persistent 
current in a zigzag carbon nanotube comes only from the individual zigzag 
channels, not from the armchair paths. To justify it in Fig.~\ref{yvelocity} 
we present the variation of persistent current in an armchair path as a 
function of $\phi$ for a half-filled zigzag nanotube considering $N_x=20$ 
and $N_y=7$, which clearly shows zero current for the entire range of 
$\phi$. 

\begin{figure}[ht]
{\centering \resizebox*{8.2cm}{6cm}{\includegraphics{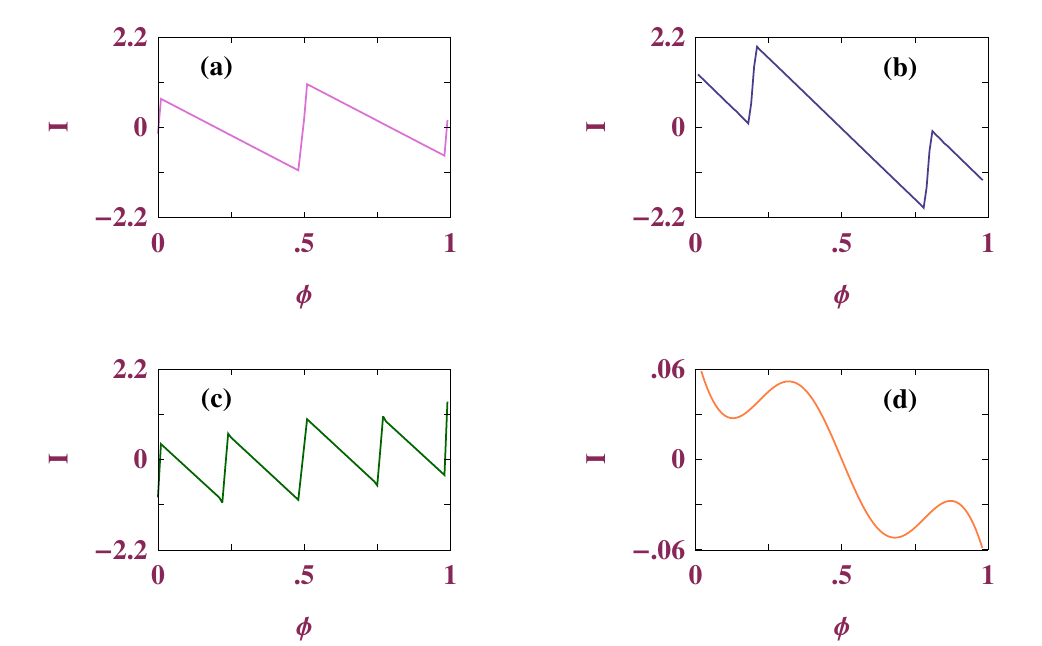}}\par}
\caption{(Color online). Current-flux characteristics of a zigzag nanotube 
with $N_x=14$ and $N_y=6$, where (a), (b), (c) and (d) correspond to 
$N_e=10$, $20$, $30$ and $82$, respectively. $U$ is fixed at $1$.}
\label{filling}
\end{figure}

In Fig.~\ref{energy} we have already observed a few almost flat energy 
levels. These energy levels contribute a very little to the persistent 
current while, on the other hand, the energy levels with larger slopes 
provide large persistent current. Also, at the minima or maxima points 
persistent current becomes zero which is quite obvious since the current 
is obtained by taking the first order derivative of the eigen energy with 
respect to flux $\phi$ (Eq.~\ref{deri}). This peculiar nature of the 
energy levels invokes the {\em current amplitude to become filling 
dependent} and we elaborate it in this section. To clarify this feature 
in Fig.~\ref{filling} the current-flux characteristics of a zigzag nanotube, 
with $N_x=14$, $N_y=6$ and $U=1$, is depicted where four different figures 
correspond to the four different cases of electron fillings, $N_e=10$, $20$, 
$30$, and $82$. In all these cases, persistent current varies periodically 
with flux $\phi$, exhibiting $\phi_0$ flux-quantum periodicity. Also, we 
observe that there are multiple kinks in the current profiles at different 
values of $\phi$ when the number of electrons remains lower than that 
required for half-filling of the band. These kinks are associated with the 
multiple crossings of energy levels, as shown in Figs.~\ref{filling}(a)-(c) 
where the number of electrons are respectively, $10$, $20$, and $30$.
This is quite analogous to the feature of persistent current observed in 
conventional multi-channel mesoscopic cylinders. The behavior of persistent 
current gets significantly modified when the nanotube becomes half-filled 
or nearly half-filled. As illustrative example Fig.~\ref{filling}(d) is 
depicted. As we have already shown a few results for half-filled band case,
here we choose $N_e=82$ i.e., the nanotube is very near to the half-filled 
band condition. For this choice of parameter values it is examined that all 
the kinks disappear making the variation of persistent current quite 
smoother. This is analogous to the behavior of persistent current observed 
in traditional single-channel mesoscopic rings. For the cases when the 
nanotube is far away from half-filling, current amplitudes are quite 
comparable to each other (see Figs.~\ref{filling}(a)-(c)). On the other 
hand, when the tube is nearly half-filled current amplitude remarkably gets 
suppressed and this suppression is very much clear  
from Fig.~\ref{filling}(d). The reason behind this enormous reduction of 
current amplitude can be understood clearly when we look at the $E$-$\phi$ 
characteristics of Fig.~\ref{energy}. At half-filling or very close to 
half-filling, the top most filled energy level lies in the nearly flat 
region i.e., around $E=0$ (see Fig.~\ref{energy}(a)) and it contributes a 
little to the current. Moreover, when $U\ne0$, there is gap in the midband 
region. Now, for a particular filling we find the net persistent current 
taking the sum of individual contributions from the lowest filled energy 
levels, and, in this process only the contribution which comes from the 
highest occupied energy level finally survives and the rest disappear due 
to their mutual cancellations. It leads to the enormous reduction of 
persistent current amplitude in the half-filled or nearly half-filled band
case. This feature is independent of the size of the nanotube. 

Before we end this section we can emphasize that, the current amplitude in 
a zigzag nanotube is highly sensitive to the electron filling and this 
phenomenon can be utilized {\em in designing a high conducting to a low 
conducting switching operation and vice versa.}
 
\section{An ordered-disordered separated nanotube in presence of a magnetic
flux}

This section illustrates the behavior of persistent current in an 
ordered-disordered separated mesoscopic cylinder.  Persistent current in 
conventional disordered systems have already been reported by many physicists 
and Anderson type localization~\cite{anderson} due to disorder is not a new 
one. But, recent breakthroughs in the growth of semiconductor nanowires have 
opened up great opportunities to revolutionize technologies in nanoscale 
electronics~\cite{zhong}. Recent experiments on nanowires have already 
yielded some results in contrast to the phenomenon of strong localization 
due to doping. For instance, Cui {\it et al.}~\cite{cui} reported that the 
carrier mobility in boron-doped and phoporus-doped silicon nanowires under 
low dopant concentration is extremely low compared to bulk silicon but the 
conductance becomes diffusive and about five orders of magnitude larger 
with heavy doping. This motivates us to investigate whether these kinds 
of shell doped nanowires give rise to new features in the context of 
persistent current or not. 

\subsection{The model}

\begin{figure}[ht]
{\centering \resizebox*{3.9cm}{5.8cm}{\includegraphics{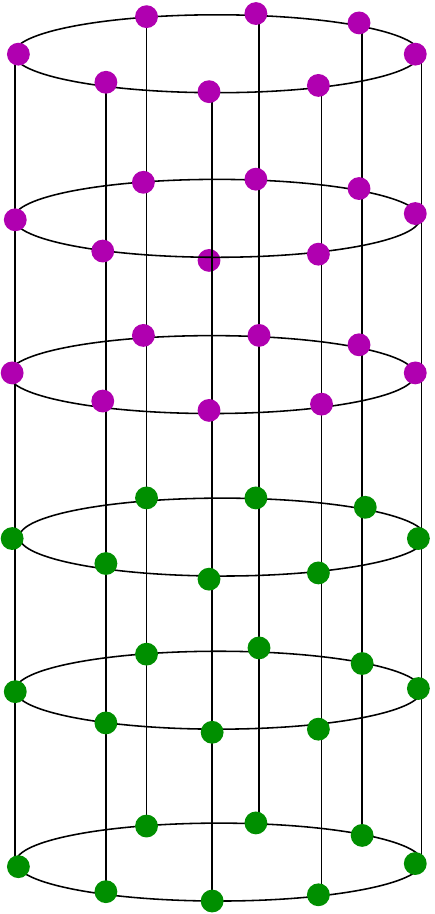}}\par}
\caption{(Color online). Schematic diagram of a mesoscopic cylinder having 
$6$ rings where each ring contains $8$ atomic sites.} 
\label{cylin}
\end{figure}

The schematic diagram of the model quantum ordered-disordered separated
nanotube is illustrated in Fig.~\ref{cylin}, where $M$ co-axial rings are 
vertically attached and each ring contains $N$ atomic sites. The cylinder 
is subjected to an AB flux $\phi$. For a comparative study we will explore 
the behavior of persistent currents in a fully disordered and an 
ordered-disordered separated cylinder simultaneously~\cite{we2}. In order 
to get an 
ordered-disordered separated cylinder disorders are introduced into $M/2$ 
number of rings of the lower half of the system (green sites) whereas the 
remaining part (magenta sites) becomes the ordered one. On the other hand, 
full disordered system is obtained by putting random disorder in all the 
atomic sites. A non-interacting single-band tight-binding (TB) Hamiltonian 
is used to describe the model quantum system which reads as,
\begin{eqnarray}
\mbox{\boldmath $H$}&=&\sum_{m,n} \left[\epsilon_{m,n} 
\mbox{\boldmath $c$}^{\dag}_{m,n} \mbox{\boldmath $c$}_{m,n} 
+ t_r (e^{i \theta_r} \mbox{\boldmath $c$}^{\dag}_{m,n} 
\mbox{\boldmath $c$}_{m,n+1}+h.c.) \right. \nonumber \\ 
& & ~~~~~~~~~~~~\left. + t_v (e^{i \theta_v} \mbox{\boldmath $c$}^{\dag}_{m,n} 
\mbox{\boldmath $c$}_{m+1,n}+h.c.)\right]
\label{hr_cyl}
\end{eqnarray}
where, $\epsilon_{m,n}$ is the on-site energy. For the impurity sites, 
the site energies $\epsilon_{m,n}$ are selected randomly from a `Box'
distribution function of width $W=1$. $t_r$ and $t_v$ refer to the 
intrachannel and interchannel nearest-neighbor couplings, respectively. 
Here, ($m$,$n$) is the 
co-ordinate of a lattice point where, $m$ and $n$ run from $1$ to $M$ and 
$N$, respectively. To incorporate the effect of magnetic flux we consider 
$\theta_r=2 \pi \phi /N$ with the threading magnetic flux $\phi$ measured 
in unit of the elementary flux-quantum $\phi_0$. The phase factor along 
the vertical direction $\theta_v$ is taken as zero. 
$\mbox{\boldmath $c$}^{\dag}_{m,n}$ ($\mbox{\boldmath $c$}_{m,n}$) is the 
creation (annihilation) operator of an electron at the site ($m,n$). At 
absolute zero temperature, the total persistent current of the cylinder is 
determined from the relation of Eq.~\ref{deri}, where $E_0$ is the ground 
state energy of the system which is obtained by taking the sum over lowest 
$N_e$ (number of electrons) energy levels. 

\subsection{Band structure}

\begin{figure}[ht]
{\centering \resizebox*{5.9cm}{3.8cm}
{\includegraphics{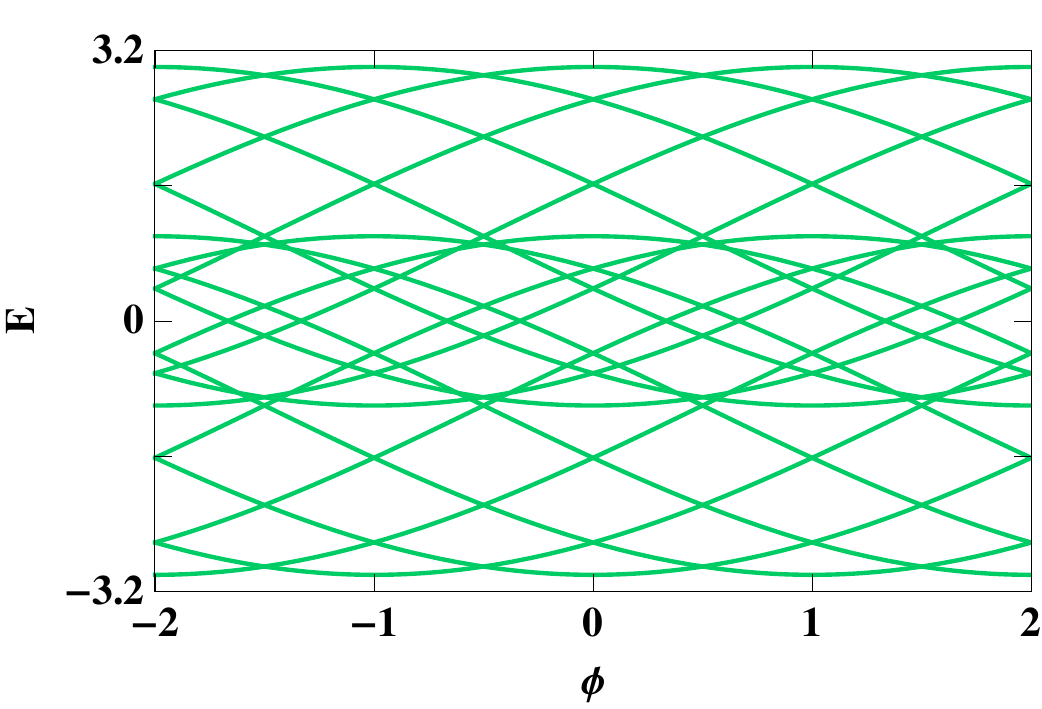}}\par}
\caption{(Color online). Energy-flux characteristics for a mesoscopic 
cylinder with $2$ rings where each ring contains $10$ atomic sites.}
\label{band_cylin}
\end{figure}

\begin{figure}[ht]
{\centering \resizebox*{5.9cm}{3.8cm}
{\includegraphics{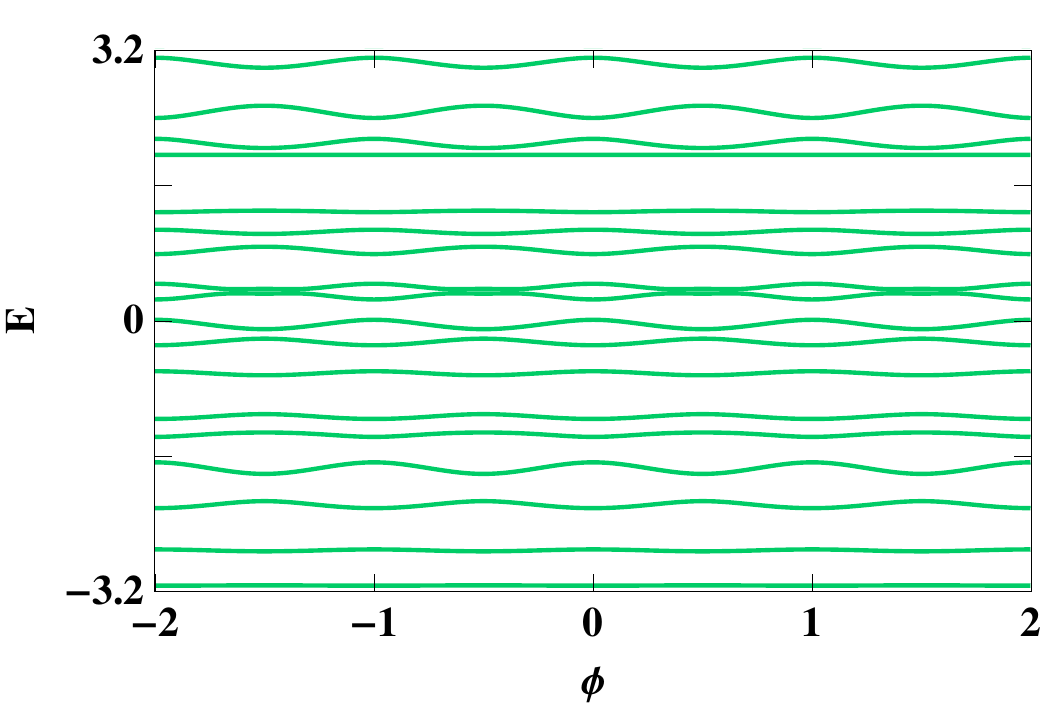}}\par}
\caption{(Color online). Energy-flux characteristics for a mesoscopic
cylinder with the same parameter values mentioned in Fig.~\ref{band_cylin}
in presence of disorder at all atomic sites ($W=3.5$).}
\label{bandw}
\end{figure}

In this sub-section we briefly discuss the energy band spectra of a 
mesoscopic cylinder. To find an analytic form of the energy levels we 
take a very small sized cylinder having only two rings with $10$ 
sites in each ring. For this smallest possible size of the cylinder, 
expression of the energy levels takes the form,
\begin{equation}
E=\pm t_v + 2 t_r Cos\left[\frac{2 \pi (n+ \phi)}{N_s \phi_0}\right]
\label{spectra_cylin}
\end{equation}
where, $N_s(=M$x$N)$ is the total number of atomic sites of the cylinder. 
In Fig.~\ref{band_cylin} we plot the variation of energy levels as a
function of magnetic flux $\phi$. The numerical values of the parameters 
are taken as $t_r=t_v=-1$ and $a=1$, where $a$ is the lattice parameter. 
Two energy bands consisting of $N$ number of energy levels within the 
boundary $E=-1$ to $+3$ and $-3$ to $+1$ are found to overlap in the 
region $E=-1$ to $+1$ (in unit of $t_r$). The width of the overlap region 
can be tuned by tuning
the vertical hopping strength ($t_v$). Each energy level exhibits one 
flux-quantum quantum periodicity. The variation of the energy levels of an 
ordered multi-channel system are already discussed in the previous section. 
Effect of the disorder can be attributed to the removal of degeneracy of 
the energy levels which can be realized from Fig.~\ref{bandw}. The results 
are calculated numerically for a particular disorder configuration. In 
presence of impurity gap opens up at the crossing points and width of this 
gap increases with the increase of the disorder strength making the slope 
of the energy levels smaller and smaller. This makes a remarkable change in 
magnitude of persistent current. We discuss it in the forthcoming sub-section. 

\begin{figure}[ht]
{\centering \resizebox*{8.0cm}{3.8cm}
{\includegraphics{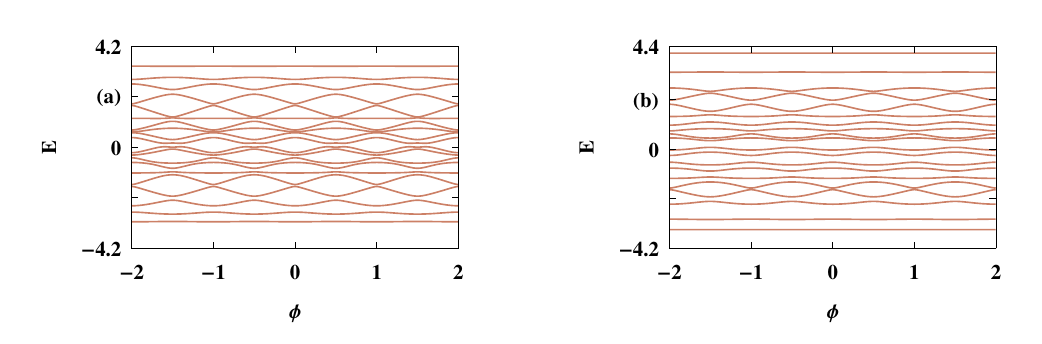}}\par}
\caption{(Color online). $E$-$\phi$ curves for an ordered-disordered 
mesoscopic cylinder with $2$ rings and $10$ atomic sites in each ring.}
\label{bandw2}
\end{figure}

Not only the full disordered cylinder here we also present the energy 
spectra of an ordered-disordered separated cylinder in Fig.~\ref{bandw2}, 
where (a) and (b) correspond to two different values of the disorder 
strengths, $W=2.5$ and $5$, respectively. Here, the nature of the variation 
of energy levels are much more different than that in the case of full 
disordered system. In order-disordered separated nanotube the energy levels 
become flatter mainly from the boundary of the spectra.

\subsection{Results and discussion}

\begin{figure}[ht]
{\centering \resizebox*{8.9cm}{4.5cm}{\includegraphics{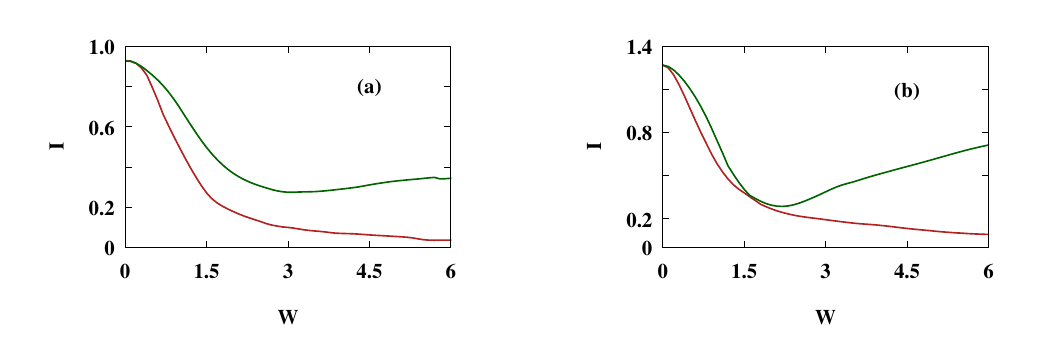}}\par}
\caption{(Color online). I-W characteristics of an ordered-disordered
separated cylinder in the half-filled band case. (a) $M=10$, $N=12$ and 
$\phi=0.4$ and (b) $M=8$, $N=10$, and $\phi=0.2$. The red and green color 
correspond to full disordered and half-disordered cylinder.}
\label{currW}
\end{figure}

In order to study the effect of disorder on persistent current in 
Fig.~\ref{currW} we plot the typical persistent current amplitude $I$ 
at a particular value of $\phi$ as a function of the disorder strength 
$W$ for an ordered-disordered separated cylinder. Figure (a) corresponds 
to the results for a half-filled cylinder consisting of $10$ rings with
$12$ sites in each ring and the magnetic flux $\phi=0.4$. On the other 
hand, figure (b) represents the results for a  half-filled cylinder
having $8$ rings ($10$ atomic sites in each ring) in presence of magnetic 
flux $\phi=0.2$. Here we compute the root mean square of the current 
amplitude taking the average over $30$ random disordered configurations. 
We observe that for a full disordered cylinder the current amplitude 
decreases with the rise of impurity strength $W$ (red curve), while in the
case of ordered-disordered separated cylinder the current amplitude initially 
decreases upto a certain value of $W$ after which it slowly increases with 
the rise of the impurity strength (green curve). This phenomenon can be 
explained as follows. In the presence of disorder current amplitude decreases 
due to the localization of the energy levels which is the so-called Anderson 
localization. The more impurity strength results the more reduction in 
current amplitude. This is true only for a full disordered system. But, for 
the ordered-disordered separated system there is a different physical picture. 
As we gradually tune the disorder strength towards the higher value, the 
ordered and disordered regions get decoupled from each other and after a 
critical limit of $W$, (say $W_c$), the cylinder behaves as 
composed of two completely decoupled regions. At this situation only the 
ordered region contributes to the current, and thus, the current amplitude
gets enhanced with the disorder strength $W$. The critical value $W_c$ 
depends on the system size and other parameter values like magnetic flux. 
As for example, it is $3$ and $2.5$ for two different system sizes as shown 
in Figs.~\ref{currW}(a) and (b), respectively. 
  
To summarize, in this section we have explored the behavior of persistent 
current in a disordered and an ordered-disordered separated cylinder 
in presence a magnetic flux. Most interestingly, we see that the current 
amplitude shows an anomalous behavior with the increase of impurity strength 
for an ordered-disordered separated cylinder in contrast to the completely 
disordered one. This study may be helpful for exploring 
localization-delocalization transition in shell-doped nanotubes.

\section{A binary alloy ring without external electrodes}

In this section we undertake an analysis of the band structure and 
persistent current in an isolated binary alloy ring (no source and
drain electrodes) enclosing a magnetic flux $\phi$~\cite{we3}  

\subsection{The model}

\begin{figure}[ht]
{\centering \resizebox*{5.5cm}{4.2cm}{\includegraphics{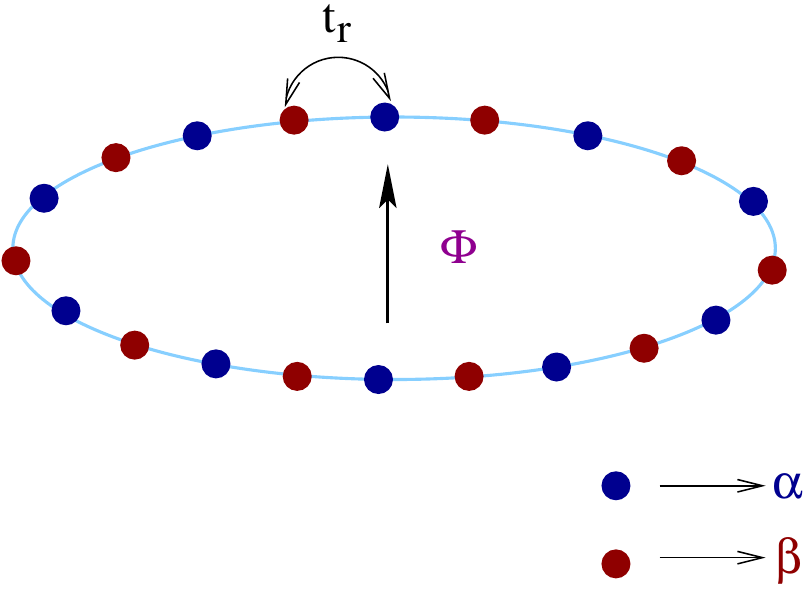}}\par}
\caption{(Color online). A binary alloy ring, threaded by a magnetic flux
$\phi$, is composed of two different types of atomic sites, viz, $\alpha$
and $\beta$ those are represented by filled blue and red circles,
respectively.}
\label{ring1}
\end{figure}

Let us concentrate on the simplest model of a binary alloy ring as shown in 
Fig.~\ref{ring1}, where the ring consists of two different types of atoms
placed alternately in a regular pattern. They are characterized by two 
different on-site potential energies, namely, $\alpha$ and $\beta$. The 
ring is subjected to an AB flux $\phi$. Within a non-interacting 
single-band TB framework we illustrate the model of binary alloy ring and 
the TB Hamiltonian reads,
\begin{equation}
\mbox{\boldmath $H$}_{R}=\sum_l (\epsilon_l \mbox{\boldmath $c$}^{\dag}_l 
\mbox{\boldmath $c$}_l + t_r e^{i \theta} \mbox{\boldmath $c$}^{\dag}_l
\mbox{\boldmath $c$}_{l+1} + t_r e^{-i \theta} 
\mbox{\boldmath $c$}^{\dag}_{l+1} \mbox{\boldmath $c$}_{l})
\label{hr}
\end{equation}
where, the on-site energy, $\epsilon_l$ takes two values $\epsilon_{\alpha}$
and $\epsilon_{\beta}$ corresponding to two different sites $\alpha$ and 
$\beta$, respectively. $t_r$ is the nearest-neighbor hopping integral. The 
phase factor $\theta=2\pi\phi/N$ of the Hamiltonian takes an account of the 
effect of the magnetic flux $\phi$ threaded by the ring which is measured in
unit of the elementary flux-quantum $\phi_0$. 
$\mbox{\boldmath $c$}_l^{\dagger}$ ($\mbox{\boldmath $c$}_l$) is the creation 
(annihilation) operator of an electron at the site $l$. Here, $l$ runs from 
$1$ to $N$, where $N$ is the total number of sites in the binary ring.

\subsection{Energy spectrum}

\begin{figure}[ht]
{\centering \resizebox*{6.5cm}{4.6cm}{\includegraphics{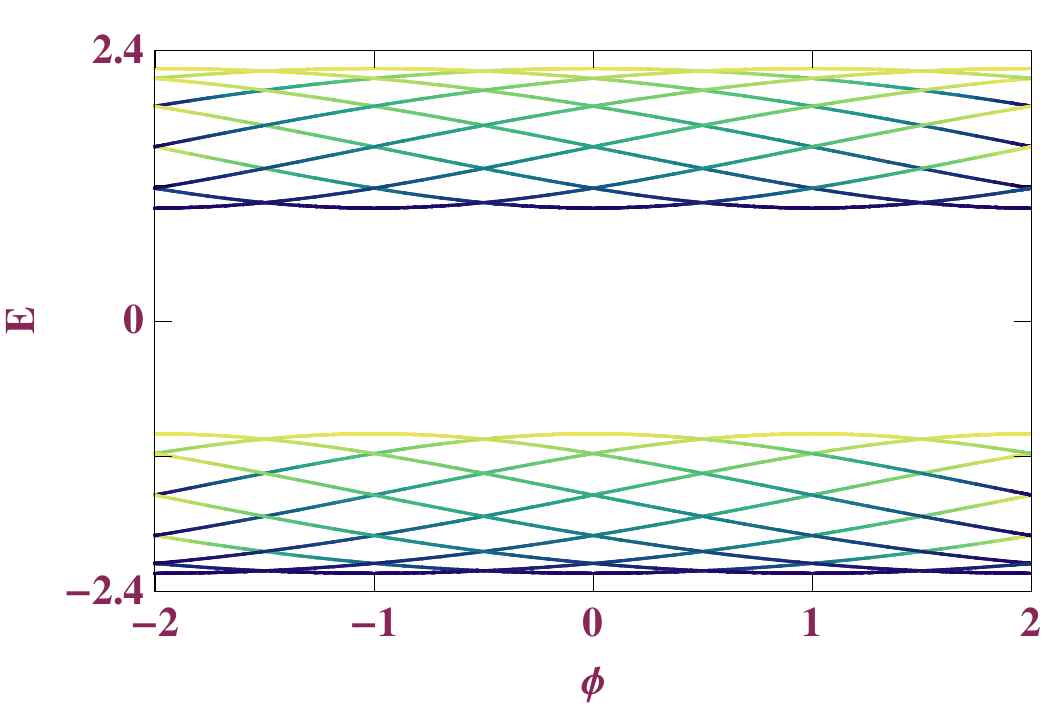}}\par}
\caption{(Color online). Energy-flux characteristics of an ordered binary
alloy ring ($N=40$) considering $\epsilon_{\alpha}=-\epsilon_{\beta}=1$
and $t_r=1$. $\phi_0$ is set at $1$.}
\label{band_ring}
\end{figure}

Before addressing the main points i.e., the characteristic features of 
persistent current in an ordered binary alloy ring, let us have an idea 
about the energy band structure of the system. The analytical expression of 
energy dispersion 
relation for the ordered binary alloy ring is as follows,
\begin{equation}
E=\frac{\epsilon_{\alpha}+\epsilon_{\beta}}{2} \pm
\sqrt{\left(\frac{\epsilon_{\alpha}-\epsilon_{\beta}}{2}\right)^2+
4\, t_r^2 \, cos^2{\left(ka\right)}}
\label{en_k1}
\end{equation}
where, $a$ is the lattice spacing and $k$ is the wave vector. The periodic 
boundary condition of the ring sets the quantized values of $k$ in presence
of the AB flux $\phi$ as,
\begin{equation}
k=\frac{2 \pi}{N a} \left(n+\frac{\phi}{\phi_0}\right)
\end{equation}
where, $n$ is an integer and it is restricted within the range: $-N/2 \le
n< N/2$. Throughout our manuscript we consider 
$\epsilon_{\alpha}=-\epsilon_{\beta}=\epsilon$ and Eq.~\ref{en_k1} is 
modified according to this condition as,
\begin{equation}
E= \pm \sqrt{\epsilon^2+ 4\, t_r^2 \, cos^2{\left(ka\right)}}
\label{en_k2}.
\end{equation}
In Fig.~\ref{band_ring} we plot the energy levels as a function of flux 
$\phi$, obtained from Eq.~\ref{en_k2}, for a $40$-site binary alloy ring 
considering $\epsilon=1$ and $t_r=1$. Looking at Fig.~\ref{band_ring}, two 
different sets of energy levels are noticed to form two quasi-bands 
separated by a finite energy gap. This gap, on the other hand, is tunable 
by the parameter values describing the TB Hamiltonian Eq.~\ref{hr}. The 
origin of two different sets of energy levels is also clearly understood 
from Eq.~\ref{en_k2}. The energy levels of Fig.~\ref{band_ring} have either 
a maximum or a minimum at half-integer or integer multiples of flux-quantum
and it results vanishing nature of persistent current at these specific 
values of $\phi$, since the current is obtained by taking the first order 
derivative of energy $E(\phi)$ with
respect to flux $\phi$. All these energy levels vary periodically providing 
$\phi_0$ flux-quantum periodicity, $\phi_0$ being $1$ in our chosen unit 
($c=h=e=1$).

\subsection{Persistent current}

\begin{figure}[ht]
{\centering \resizebox*{7.5cm}{7cm}{\includegraphics{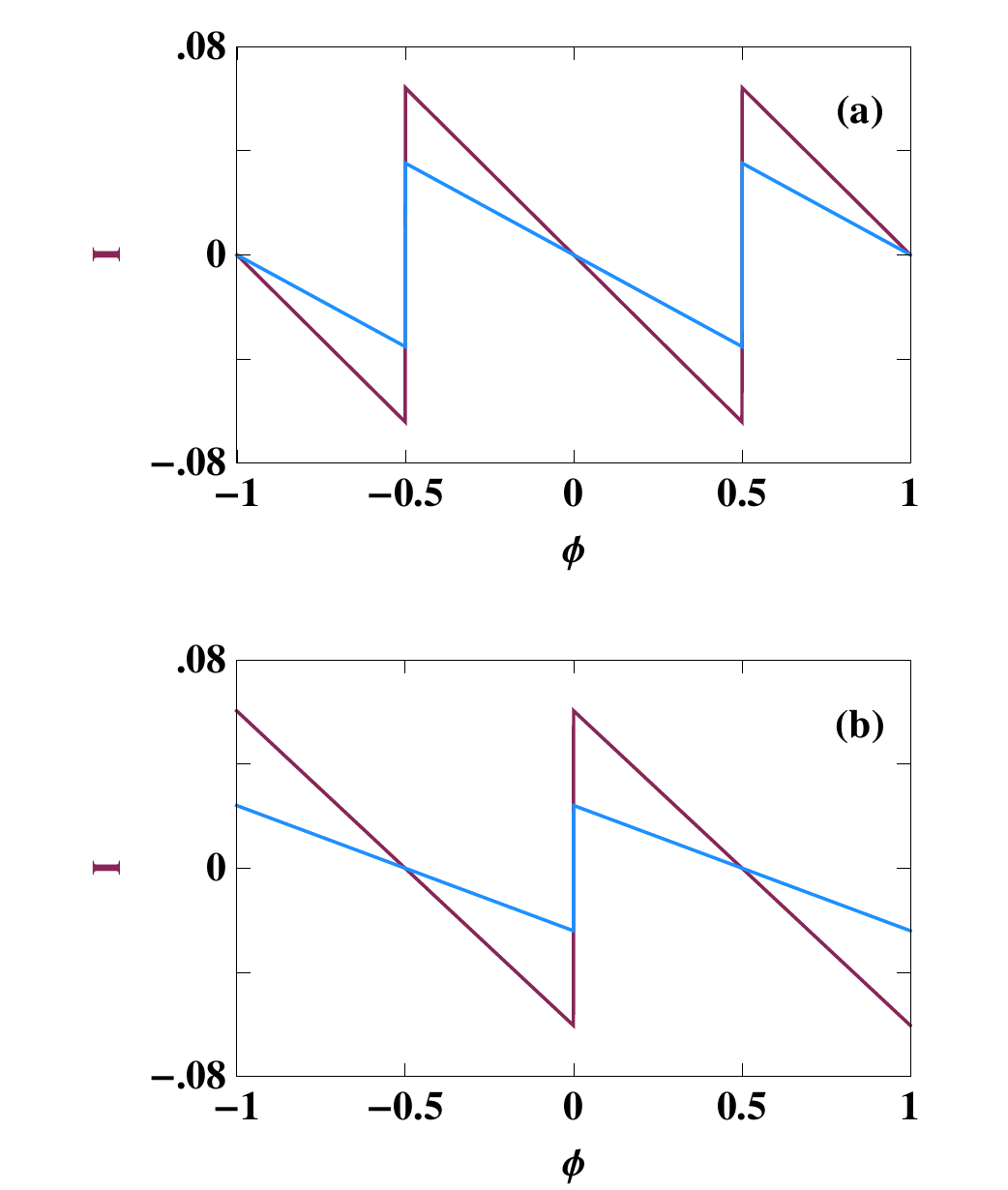}}\par}
\caption{(Color online). Current-flux characteristics of an ordered binary
alloy ring ($N=120$) considering $\epsilon_{\alpha}=-\epsilon_{\beta}=1$
and $t_r=1$. The blue and purple colors in (a) correspond to $N_e=15$
and $35$, while in (b) they correspond to $N_e=10$ and $30$, respectively.
The lattice spacing $a$ is set at $1$ and we choose $\phi_0=1$.}
\label{oddeven}
\end{figure}

Our task of calculating persistent current for individual energy 
eigenstates is now easier as we know the energy eigenvalues of the ring 
as a function of flux $\phi$. It is simply the first order derivative of 
energy with respect to flux $\phi$. Therefore, for an $n$-th energy 
eigenstate we can write the expression for the current as,
\begin{equation}
I_n=\pm \left(\frac{4 \pi t_r^2}{N a \phi_0} \right)
\frac{sin{\left[\frac{4 \pi}{N a}\left(n+\phi/\phi_0\right)\right]}}
{\sqrt{1+ 4 t_r^2 cos^2{\left[ \frac{2 \pi}{N a} \left(n+\phi/\phi_0\right)
\right]}}}
\end{equation}
where, $+$ve or $-$ve sign appears in the current expression depending on 
the choice of $n$ i.e., in which sub-band the energy level exists. To get 
total persistent current $I$ for a particular filling $N_e$, we take the 
sum of individual contributions from the lowest $N_e$ energy eigenstates 
as we do our calculations at absolute zero temperature. The expression is,
\begin{equation}
I=\sum_{n=1}^{N_e} I_n
\end{equation}
Following this way, we plot the variation of persistent current for an 
ordered $120$-site binary alloy ring in Fig.~\ref{oddeven}. The values of 
the parameters considered in this figure are $\epsilon=1$ and $t_r=1$. In
both figures, Fig.~\ref{oddeven}(a) and (b), the current profiles show 
saw-tooth like variation as a function of flux $\phi$, similar to that of 
traditional single-channel mesoscopic rings~\cite{san7}. The purple and 
blue colors correspond to two different fillings of the band. For odd 
number of electrons the results are shown in Fig.~\ref{oddeven}(a) while, 
in Fig.~\ref{oddeven}(b) the results are given for even number of 
electrons. The sharp transitions at half-integer (for odd $N_e$) or integer 
(for even $N_e$) multiples of flux-quantum ($\phi_0$) in persistent current 
appears due to the crossing of energy levels at these respective values 
of $\phi$. Quite interestingly, we also examine that the current shows 
always diamagnetic response irrespective of the filling factor.

\section{A binary alloy ring with external electrodes}

Upto now we have discussed the basic features of persistent current in 
different isolated mesoscopic systems. In this section we extend our 
analysis of persistent current to an open system where a binary alloy 
ring~\cite{we2} is clamped between two semi-infinite one-dimensional 
($1$D) electrodes. The behavior of persistent current in presence of 
transport current will also be analyzed.

\subsection{The model}

\begin{figure}[ht]
{\centering \resizebox*{8.0cm}{4.8cm}{\includegraphics{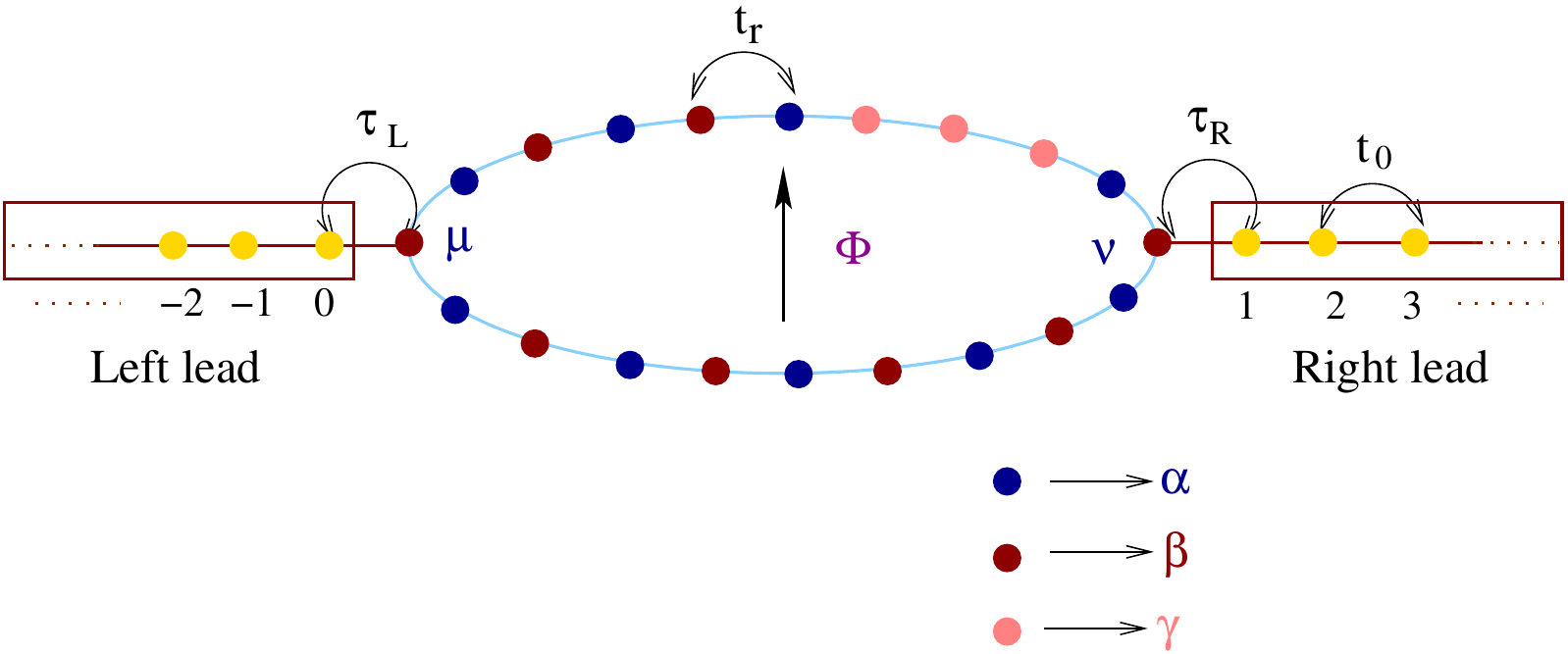}}\par}
\caption{(Color online). A binary alloy ring in presence of some identical
foreign atoms (labeled as $\gamma$ sites), threaded by an AB flux $\phi$,
is attached to two semi-infinite one-dimensional metallic leads. $\mu$ and
$\nu$ are the connecting sites.}
\label{ring2}
\end{figure}

Let us start by referring to Fig.~\ref{ring2} where a binary alloy ring, 
threaded by a magnetic flux $\phi$, is attached to two semi-infinite 
one-dimensional metallic electrodes, namely, left-lead and right-lead, 
via two atomic sites labeled as $\mu$ and $\nu$. The total numbers of 
identical pairs of $\alpha$-$\beta$ sites in the binary ring is $N_1$. 
Now we incorporate some foreign atoms denoted by $\gamma$ having on-site 
potentials $\epsilon_\gamma$ in any one of the two arms of the ring. 
Here, we choose the upper arm in this purpose. There are $N_2$ identical 
foreign atomic sites, embedded together in a small portion of the ring. 
These $\gamma$ sites are often referred to as impurity sites in the present 
manuscript. We enumerate the atomic sites of the two side-attached leads in 
a particular way, as shown in Fig.~\ref{ring2}. A single-band non-interacting 
TB framework is used to describe the entire system. For the full system we 
can partition the total Hamiltonian as a sum of three terms like,
\begin{equation}
\mbox {\boldmath{$H$}}=\mbox {\boldmath{$H$}}_{R}+\mbox {\boldmath{$H$}}_L
+\mbox {\boldmath{$H$}}_T
\label{part}
\end{equation}
where, $\mbox {\boldmath{$H$}}_R$, $\mbox {\boldmath{$H$}}_L$ and 
$\mbox {\boldmath{$H$}}_T$ represent the Hamiltonians for the ring, leads 
(left and right) and coupling between the ring and leads, respectively. The 
ring Hamiltonian ($H_R$) takes the form exactly similar to Eq.~\ref{hr}, 
but instead of two possible on-site potentials here $\epsilon_l$ has three 
possibilities for three different atomic sites ($\alpha$, $\beta$ and 
$\gamma$). By means of some external gate voltage $V_g$, the site energy 
$\epsilon_{\gamma}$ can be tuned and accordingly the site-energies are 
changed. Thus we express the on-site energy of a single $\gamma$ atom like, 
$\epsilon_{\gamma}=\epsilon_{\gamma}^0 + V_g$, where $\epsilon_{\gamma}^0$ 
is the site energy in absence of any external potential. The other two terms 
of Eq.~\ref{part}, $\mbox {\boldmath{$H$}}_L$ and $\mbox {\boldmath{$H$}}_T$, 
can also be written in a similar fashion as,
\begin{equation}
\mbox {\boldmath{$H$}}_L=\underbrace {t_0 \sum_{m \le 0} 
\left(\mbox {\boldmath{$b$}}_m^{\dag} \mbox {\boldmath{$b$}}_{m-1} +
h.c. \right)}_{\mbox {left lead}} +
\underbrace {t_0 \sum_{m \ge 1} \left(\mbox {\boldmath{$b$}}_m^{\dag}
\mbox {\boldmath{$b$}}_{m+1} +h.c. \right)}_{\mbox {right lead}}
\label{hlead}
\end{equation}
and,
\begin{equation}
\mbox {\boldmath{$H$}}_T=\left( \tau_L \mbox {\boldmath{$b$}}_0^{\dag} 
\mbox {\boldmath{$c$}}_{\mu} + \tau_R \mbox {\boldmath{$b$}}_1^{\dag} 
\mbox {\boldmath{$c$}}_{\nu} \right) +
h.c.
\end{equation}
where $\mbox {\boldmath{$b$}}_m^{\dag}$ ($\mbox {\boldmath{$b$}}_m$) are the 
creation (annihilation) operator of an electron at the site $m$ of the leads 
and $t_0$ represents the nearest-neighbor hopping strength within these 
leads. We set the site energy for the identical sites in the leads to zero 
and due to this reason in Eq.~\ref{hlead} we have omitted the site-energy 
terms. Here, $\tau_L$ is the coupling strength between the left lead and 
the ring, while it is $\tau_R$ for the other case.

\subsection {Wave-guide theory}

To find transmission probability across the ring and also to calculate
persistent current in such an open ring geometry we adopt the wave-guide
theory~\cite{xia,xiong}. In the present sub-section we describe the 
formulation very briefly.

We begin with the Scr\"{o}dinger equation $H |\psi \rangle = E |\psi\rangle$, 
where $|\psi \rangle$ is the stationary wave function of the entire system. 
In the Wannier basis it ($|\psi \rangle$) can be expressed as,
\begin{equation}
|\psi \rangle= \underbrace{\sum_{m \le 0} B_m |m \rangle}_{\mbox{left lead}}
+ \underbrace{\sum_{m \ge 1} B_m |m \rangle}_{\mbox{right lead}} +
\underbrace{\sum_{l} C_l |l \rangle}_{\mbox{ring}}
\label{shy}
\end{equation}
where, the co-efficients $B_m$ and $C_l$ correspond to the probability
amplitudes in the respective sites. Keeping in mind the periodicity of the
ordered binary ring we write the wave functions associated with the electrons
as a plane wave and the wave amplitudes in the left and right leads are,
\begin{equation}
B_m=e^{ikm}+r e^{-ikm}, \hskip 0.3cm {\mbox{for}} \hskip 0.3cm m \le 0
\label{equ10}
\end{equation}
and
\begin{equation}
B_m=t e^{ikm}, \hskip 0.3cm {\mbox{for}} \hskip 0.3cm m \ge 1
\label{equ11}
\end{equation}
where, $r$ and $t$ are the reflection and transmission amplitudes,
respectively. $k$ is the wave number and it is related to the energy
$E$ of the incident electron by the expression $E=2t_0 \cos{k}$. The
lattice spacing $a$ is set equal to $1$.

In order to find out the transmission amplitude $t$, we have to solve the
following set of coupled linear equations.
\begin{eqnarray}
E B_0 &=& t_0 B_{-1} + \tau_L C_{\mu} \nonumber \\
(E-\epsilon_l) C_l &=&t _r e^{i \phi} C_{l+1}+t_r e^{-i \phi} C_{l-1}+
\tau_L B_0 \delta_{l,\mu} \nonumber \\
~& &~~~~~~~~~~~~+ \tau_R B_1 \delta_{l,\nu} \nonumber \\
E B_1&= &t_0 B_2 + \tau_R C_{\nu}
\label{seteqns}
\end{eqnarray}
where, the co-efficients $B_0$, $B_{-1}$, $B_1$ and $B_2$ can be easily
expressed in terms of $r$ and $t$ by using Eqs.~\ref{equ10} and \ref{equ11},
and they are in the form:
\begin{eqnarray}
B_0 &=& 1+r \nonumber \\
B_{-1}&=& B_0 e^{ik}-2i \sin{k} \nonumber \\
B_1 &=& t e^{ik} \nonumber \\
B_2 &=& t e^{2ik}
\end{eqnarray}
Thus, for a particular value of $E$ we can easily solve the set of linear
equations and find the value of $t$. Finally, the transmission probability
across the ring becomes
\begin{equation}
T(E)=|t|^2.
\end{equation}
Now, to compute the persistent current between any two neighboring sites in 
the binary ring we use the following relation,
\begin{equation}
I_{l,l+1}=\frac{2 e t_r}{N \hbar} {\mbox{Im}}\left( C_l^{\dag} C_{l+1}
e^{-i \phi}
\right).
\end{equation}

\subsection{Transmission and average density of states}

Throughout our calculations we set $\epsilon=1$, $\epsilon_{\gamma}^0=0$, 
$t_r=1$, $\epsilon_0=0$ and $t_0=2$. The energy scale is measured in unit 
of $t_r$.

\begin{figure}[ht]
{\centering \resizebox*{6.8cm}{10.6cm}
{\includegraphics{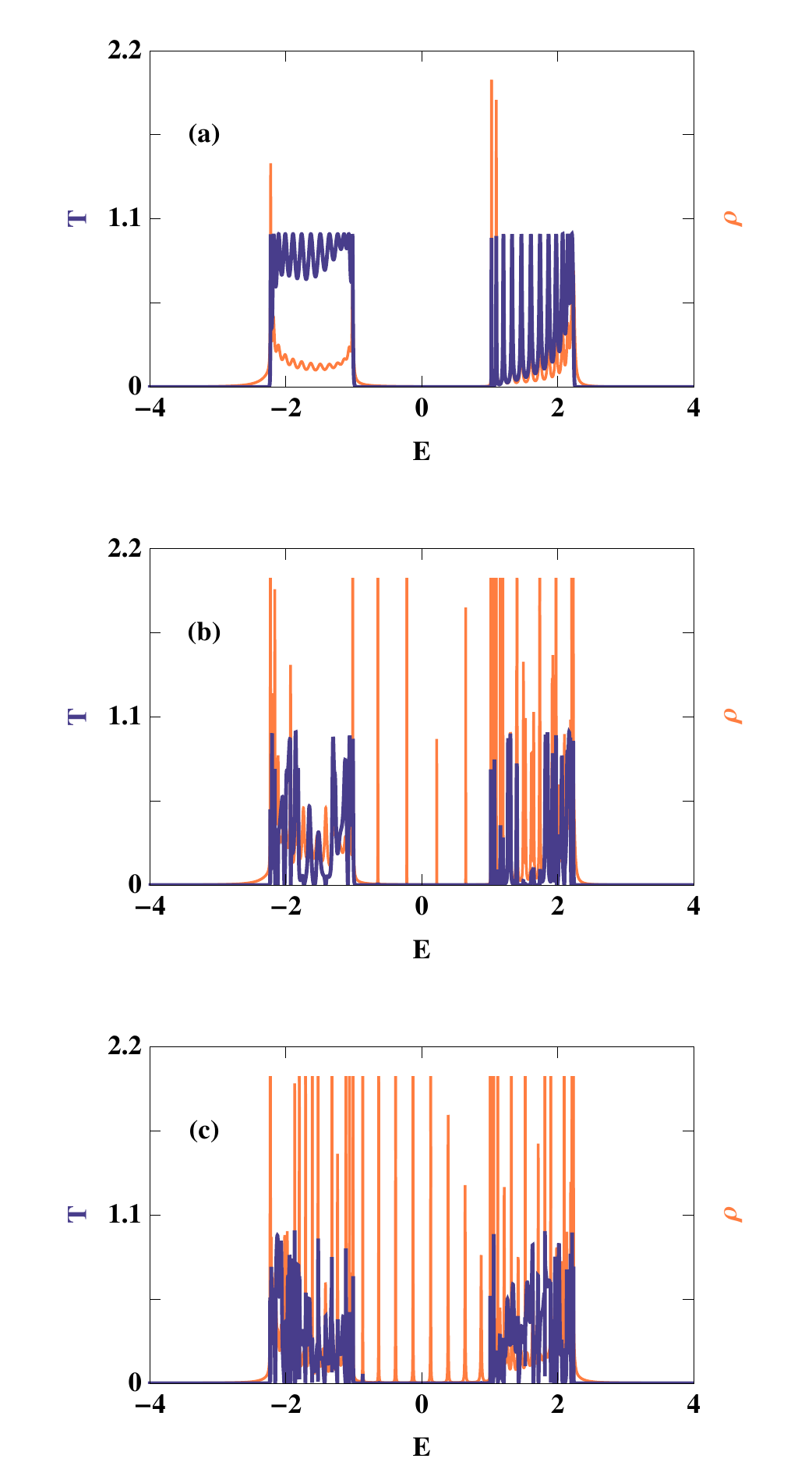}}\par}
\caption{(Color online). Transmission probability (orange color) and average
density of states (dark-blue color) for some typical binary alloy rings with
fixed number of $\alpha$-$\beta$ pairs ($N_1=28$), but different values of
impurity sites $N_2$, where (a) $N_2=0$, (b) $N_2=12$ and (c) $N_2=22$.
For these three cases we set $\mu=1$ and choose $\nu=29$, $35$ and $40$,
respectively. Other parameters are: $\epsilon_{\gamma}=0$, $\tau_L=\tau_R=1$
and $\phi=0$.}
\label{spectrum}
\end{figure}

Two-terminal transmission probability $T$ (orange color) as a function of 
injecting electron energy $E$ for some typical binary alloy rings 
considering different number of impurity sites is displayed in 
Fig.~\ref{spectrum} where (a), (b) and (c) correspond to three different
numbers of impurity atoms. Figure (a) represents the transmission spectrum
of the binary alloy ring with $28$ pairs of $\alpha$-$\beta$ atoms but 
without any foreign impurity atoms while (b) and (c) correspond to $N_2=12$ 
and $22$, respectively, keeping $N_1$ fixed to $28$. The average density of 
states (ADOS) is also superimposed in each spectrum. In all these cases the 
magnetic flux $\phi$ is set equal to zero. The results are quite interesting. 
In absence of any impurity site the transmission spectrum is characterized by
two bands separated by a finite gap and the gap between these two bands are 
also tunable by the parameter values describing the system. Looking back at 
Fig.~\ref{band_ring} the band splitting for the ordered binary alloy ring is 
easily understood and we predict that the spectrum is just the finger print 
of the system energy levels. This transmission spectrum exactly overlaps 
with the ADOS profile ($\rho$-$E$ spectrum) which ensures that electronic 
transmission takes place through all the energy eigenstates of the binary 
alloy ring and they are extended in nature.
The situation becomes really interesting when some additional impurities 
($\gamma$ sites) are introduced in any part of the binary alloy ring. Due 
to the inclusion of such atomic sites some energy levels appear within 
the band of extended regions those are no longer extended, but they are 
almost quasi-localized and do not contribute to the electronic transmission. 
This behavior is elaborated in Figs.~\ref{spectrum}(b) and (c), where 
$T$-$E$ and $\rho$-$E$ spectra are superimposed with each other for two 
different numbers of impurity sites ($N_2$). In the band center, several 
energy levels appear and these levels do not provide any contribution to the
transmission of electrons. They are almost localized in nature. Mainly the 
states within the two bands are responsible for electron transmissions. The 
number of the almost localized energy levels increases with the increase of 
the number of the impurity sites, and for sufficiently large number of 
impurities they form a quasi-energy 
band of localized states. The location of the localized energy band 
can be shifted towards the edge of extended regions simply by tuning the 
site energy of these foreign atoms, and this can be done by means of 
applying an external gate voltage $V_g$. We utilize this feature to make 
the binary alloy ring behave like an extrinsic semiconductor, either 
$p$-type or $n$-type by tuning the Fermi level to an appropriate place. 

\begin{figure}[ht]
{\centering \resizebox*{7.5cm}{8.6cm}{\includegraphics{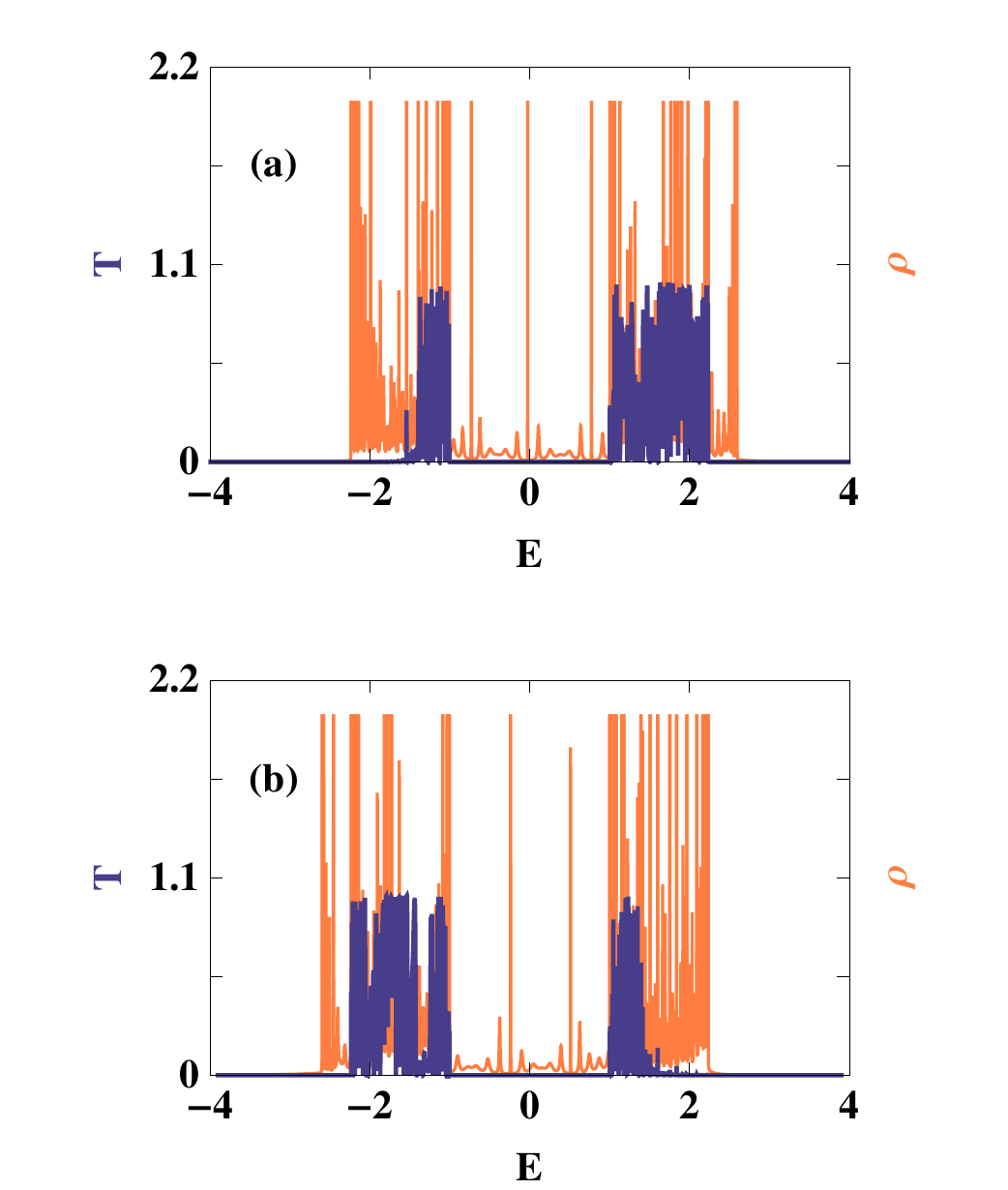}}\par}
\caption{(Color online). Transmission probability (orange color) and ADOS
(dark-blue color) as a function of energy for a binary alloy ring ($N_1=60$)
with $44$ impurity sites ($N_2=44$), where (a) and (b) correspond to
$\epsilon_{\gamma}=0.6$ and $-0.6$, respectively. Other parameters are:
$\phi=0$, $\tau_L=\tau_R=1.5$, $\mu=1$ and $\nu=83$.}
\label{pntype}
\end{figure}

We plot Fig.~\ref{pntype} to explore the semiconductor-like behavior of 
the binary ring. The variation of transmission probability and the ADOS 
are shown in this figure. Here, we consider a binary ring with $60$ 
identical pairs of $\alpha$-$\beta$ sites and $44$ number of impurity sites. 
Two different cases are exhibited in Fig.~\ref{pntype}(a) and (b) for two 
different values of the on-site potential energies of the $\gamma$ atoms. 
Figure (a) represents the transmission spectrum of the binary alloy ring 
when $\epsilon_{\gamma}$ is set at $0.6$ while the other figure 
(Fig.~\ref{pntype}(b)) corresponds to $\epsilon_{\gamma}=-0.6$. In presence
of the impurity atoms we get almost quasi-energy bands (ADOS spectra) and 
quite interestingly we observe that when $\epsilon_{\gamma}$ is fixed at 
$0.6$, a localized energy band for a wide range of energy is formed along 
the left edge of the extended region (Fig.~\ref{pntype}(a)). Now, if the 
Fermi level is set around $E=-1.7$, then many electrons in the localized 
region below the Fermi level can jump easily, even at much low temperature 
since the energy gap is almost zero, to the 
extended regions and can contribute to the current. As a result large number 
of excess electrons become available in the conduction band region which 
behave as n-type carriers. This movement however depends on the localized 
energy levels and also the available extended energy states. On the contrary, 
when the site-energies of the $\gamma$ atoms are tuned to another value, say, 
$-0.6$ an exactly opposite behavior is obtained. In this case the wide
band of localized states is formed in the right edge of the extended region
(see Fig.~\ref{pntype}(b)). Now, if the Fermi level is tuned to appropriate 
place, around $E=1.7$ in this case, then the electrons from the filled 
extended levels below the Fermi level hop to the nearly empty localized 
levels, and these electrons do not contribute anything to the current. But, 
the absence of electrons are realized by holes in the extended regions which 
can carry current and the system behaves like a $p$-type carriers. Before we 
end this discussion, we can emphasize that by setting the Fermi level in 
appropriate places our model quantum system can be tailor-made to use as a 
$p$-type or an $n$-type semiconductor.

At the end of this sub-section we would like to mention a few points. To 
establish the fact that how such a geometry can be utilized as a $p$-type 
or an $n$-type semiconductor with appropriate choice of the Fermi level 
we have taken a particular set of parameter values. As we are doing a model
calculation it is very easy to take some specific values of the parameters 
and use in our numerical calculations, but all these physical phenomena are 
exactly invariant with the change of the parameter values. Only the 
numerical values will be altered. These features are also exactly valid 
even for a non-zero value of magnetic flux $\phi$. The physical picture 
\begin{figure}[ht]
{\centering \resizebox*{8cm}{7.6cm}{\includegraphics{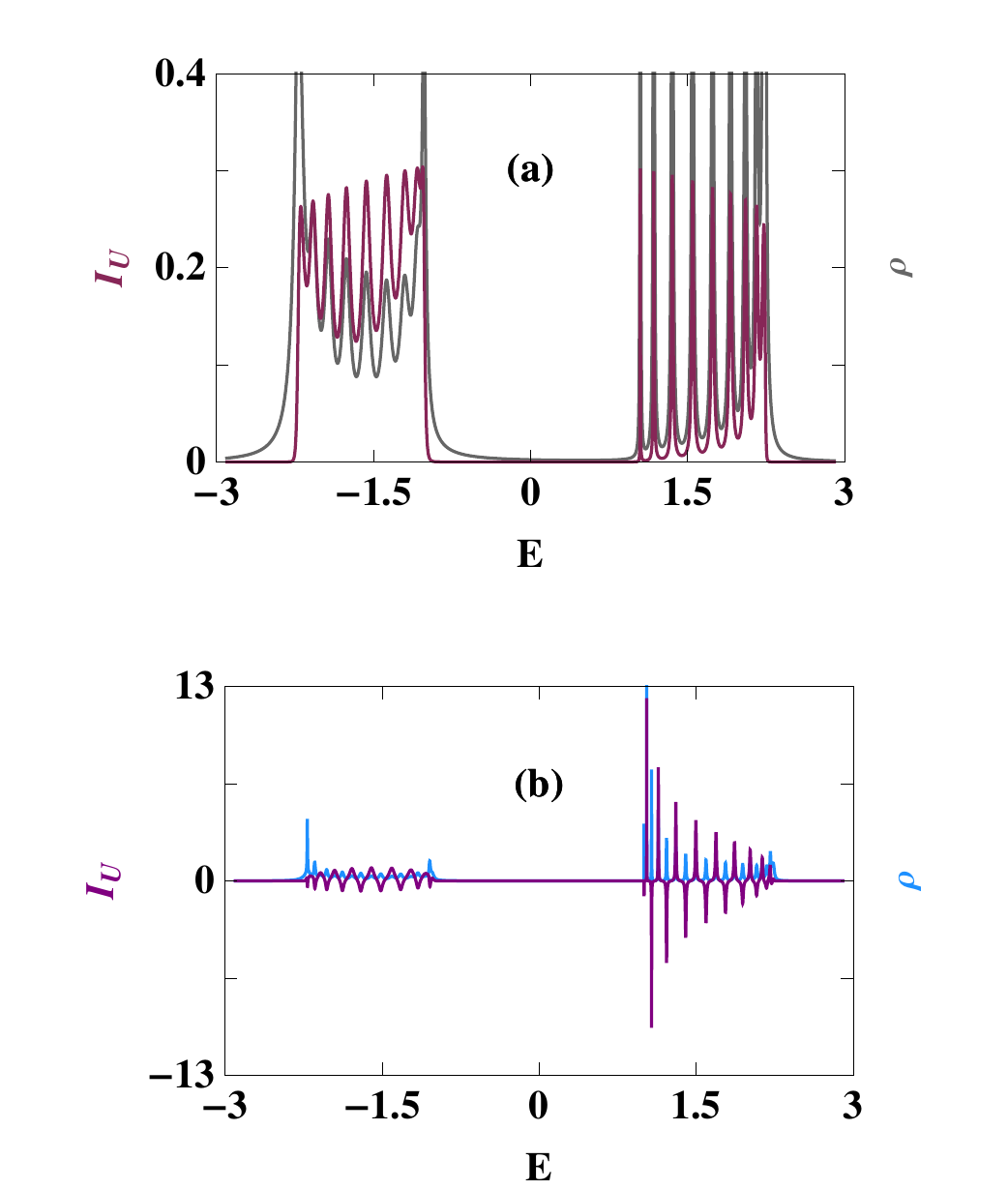}}\par}
\caption{(Color online). Persistent current ($I_U$) in the upper arm (red 
and violet color) as a function of energy $E$ for a binary alloy ring 
($N_1=20$) in the absence of impurity atoms ($N_2=0$), where (a) $\phi=0$ 
and (b) $\phi=\phi_0/4$. Other parameters are: $\tau_L=\tau_R=0.8$, 
$\mu=1$ and $\nu=21$. For this ordered ring, persistent current in the 
lower arm ($I_L$) is exactly identical to $I_U$. ADOS profile (gray and 
light-blue color) is also superimposed in each spectrum.}
\label{curr1}
\end{figure}
will be much more appealing if we consider larger rings with more impurity 
sites and it gives us the confidence to propose an experiment in this way.
In a recent work Bellucci {\em et al.}~\cite{bel1,bel2} have done a detailed
study of magneto-transport properties in quantum rings considering tunnel
barriers in the presence of magnetic field and shown how metal-to-insulator 
transition takes place in such a geometry. They have also established that 
by controlling the strength and the positions of the barriers, the energy 
shift can be done in a tunable way. This is quite analogous to our present 
study, and so, an experiment in this regard will be challenging.

\subsection{Persistent current in presence of transport current}

Now we focus on the characteristic features of persistent current in the 
binary alloy ring in presence of external bias voltage. We also study the
role of magnetic field in this context. In Fig.~\ref{curr1} upper arm 
current, $I_U$, in the upper arm of an ordered binary alloy ring as a 
function of energy $E$ is presented where (a) and (b) correspond to 
$\phi=0$ and $\phi_0/4$, respectively. Due to conservation of current at 
the junction points, current between any two sites are equal to each other. 
We refer upper arm current to the bond current between any two 
nearest-neighbor atomic sites of the upper arm. The ring is symmetrically 
coupled to the side attached leads i.e., the upper and lower arms have 
identical length, and accordingly, the
\begin{figure}[ht]
{\centering \resizebox*{8cm}{7.6cm}{\includegraphics{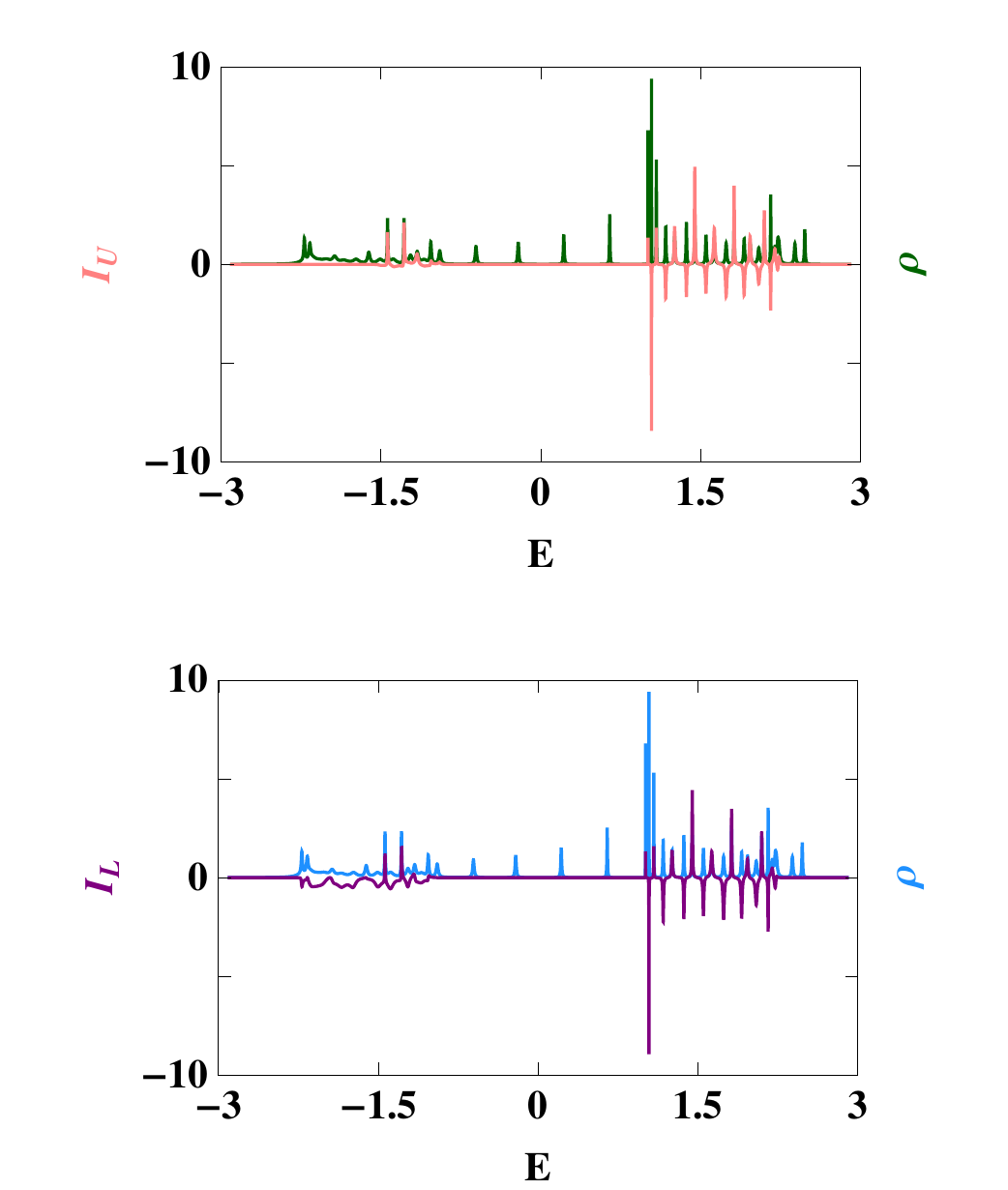}}\par}
\caption{(Color online). Persistent current as a function of energy for a
binary alloy ring ($N_1=16$) in presence of impurity sites ($N_2=12$) for
a finite value of $\phi$ ($\phi=0.3$), where (a) and (b) correspond to the
currents for the upper (pink color) and lower (violet color) arms 
respectively. Other parameters are: $\epsilon_{\gamma}=0.5$, 
$\tau_L=\tau_R=0.8$, $\mu=1$ and $\nu=23$. ADOS profile (green and 
light-blue color) is superimposed in each spectrum.}
\label{curr2}
\end{figure}
current $I_L$ in the lower arm becomes exactly identical to the current
obtained in the upper arm. Similarly, we define the lower arm current. 
Here, again the band-splitting put its mark on the current density profile. 
Finite current is available for two wide range of energies, separated by a 
finite gap, associated with the energy levels of the ring those are clearly 
visible from the ADOS profiles (green and light-blue color). The quite exact 
superposition of the 
current profile and the AVDOS profile invokes the extended natures of all
these energy eigenstates in our mind. Additionally, all the values of the
current are positive {\it i.e.} they are in the same phase ensuring the fact 
that this current is completely due the transport current between the two 
leads fixed at two different chemical potentials as there is only one driving 
force, the external bias. This phenomenon depicted in Fig.~\ref{curr1}(a) 
correspond to zero magnetic flux. As well as the magnetic field with flux 
density $\phi$ is switched on, not only both the positive and negative 
values of current appear indicating opposite phases but also they appear in 
regular alternative fashion as shown in Fig.~\ref{curr1}(b). Moreover, the 
magnitudes are also different in these two cases indicating the effect 
quantum interference of mesoscopic regime. Therefore, by measuring 
persistent current we can directly estimate the nature of the current i.e., 
\begin{figure}[ht]
{\centering \resizebox*{7cm}{4.2cm}{\includegraphics{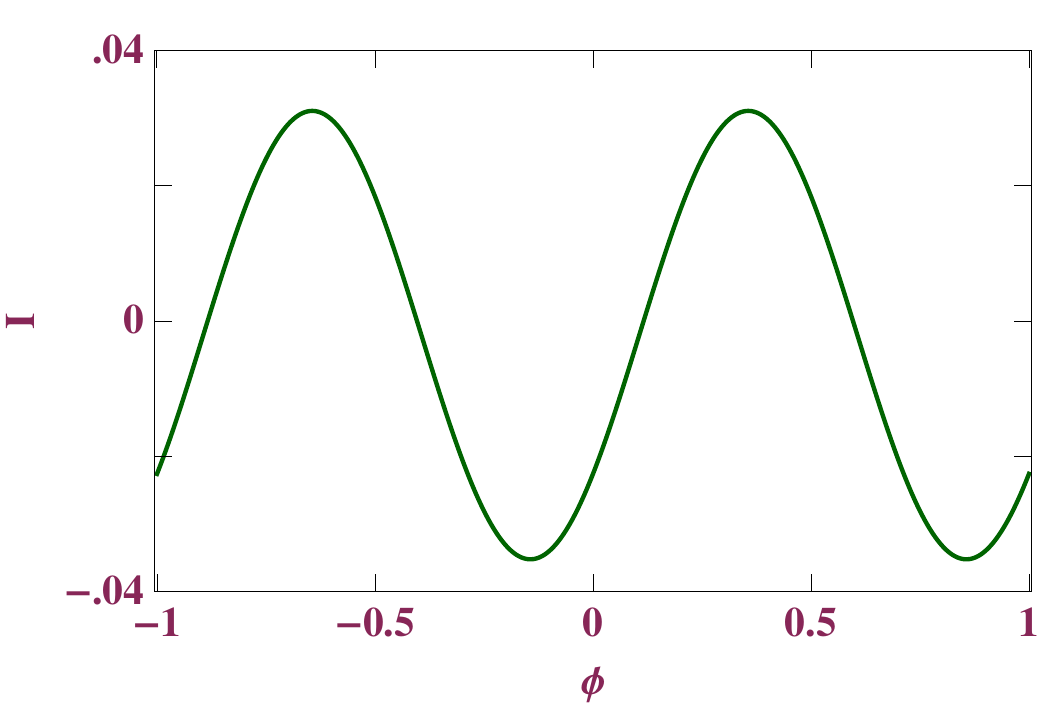}}\par}
\caption{(Color online). Persistent current as a function of magnetic flux 
for a binary alloy ring ($N_1=16$) in presence of impurity sites ($N_2=12$) 
with $V_g=0.5$ Other parameters are: $\tau_L=\tau_R=0.8$, $\mu=1$ and 
$\nu=23$.}
\label{currfi}
\end{figure}
whether it is paramagnetic or diamagnetic in nature and also predict the 
characteristics of energy eigenstates, which are somewhat interesting in 
the study of electron transport. Needless to say, magnetic field plays an 
important role in this context.

Already stated in introduction the current can persists in this kind of 
loop geometry if we set the lengths of two arms unequal or put any kind of 
static disorder in path of the electron or in any other way we can disturb 
the wave function or create any path difference between the wave functions
corresponding to different arms. In our study we incorporate impurities in 
the upper arm of the binary ring as mentioned earlier. The results are 
displayed in Fig.~\ref{curr2} where (a) and (b) represent the cases 
corresponding to the upper and lower arms of the ring, respectively. To plot 
the curve we consider a typical binary alloy ring with $N_1=16$ and $N_2=12$. 
With the inclusion of impurity sites, the symmetry between the two arms is 
broken, and therefore, the currents in the upper and lower arms are no 
longer identical to each other as shown from the spectra 
(Figs.~\ref{curr2}(a) and (b)). So, there must be a circulating current 
within the ring which is the so-called persistent current in presence of 
transport current. Unlike to the ordered binary alloy ring 
(Fig.~\ref{curr1}), here all the energy eigenstates are not extended in 
nature. Some localized energy levels appear in the band of extended energy 
states due to the presence of impurity sites in the ring. This is clearly 
visible from the spectra since for these impurity levels no current is 
available. We also check the periodicity of the persistent current. It is
shown in Fig.~\ref{currfi}. Persistent current exhibits $\phi_0$ flux-quantum 
periodicity. Thus, calculating persistent current we can emphasize the 
nature of energy eigenstates very nicely, and this idea can be utilized to 
reveal the localization properties of energy eigenstates in any complicated 
geometry.

\section{Concluding remarks}

To conclude, in the present article we have made a detailed investigation of 
magneto-transport in both closed and open mesoscopic systems. We have
started with the discussion of a zigzag nanotube pierced by a magnetic 
flux using a generalized Hartree-Fock mean field approach. Based on the 
tight-binding model we explore the effect of the Hubbard interaction on 
the energy levels of the tube. After describing different numerical 
results we have established the second quantized form to evaluate persistent 
current in individual paths of a zigzag carbon nanotube and based on this 
formulation we have also presented the numerical results for the distribution 
of persistent current in different branches of the nanotube. From the 
current-flux characteristics we have emphasized that the current amplitude 
in the zigzag nanotube is highly sensitive to the electron filling and 
this phenomenon can be utilized in designing a high conducting to a low 
conducting switching device and vice versa. Following the study of 
nanotube we have explored the persistent current in an ordered-disordered 
separated cylinder penetrated by a magnetic flux. Most interestingly, we 
have seen that the current amplitude shows an anomalous behavior with the 
increase of impurity strength. This study may be helpful for exploring 
localization-delocalization transition in shell-doped nanotubes. To 
discuss persistent current in open system i.e., system with side-attached
electrodes we have considered a binary alloy ring in presence of a magnetic 
flux $\phi$, as an illustrative example. Within a single-band non-interacting 
TB framework persistent current and band structure of an isolated ordered 
binary alloy ring have been analyzed. Then, we have explored the 
magneto-transport properties of a binary alloy ring in presence of external 
electrodes. The effect of impurities have also been addressed. Quite 
interestingly we have noticed that in the presence of some foreign atoms, 
those are not necessarily be random, in any part of the ring, some 
quasi-localized energy levels appear within the band of extended energy 
levels. The locations of these almost localized energy levels can also be 
regulated by means of some external gate voltage. This leads to a 
possibility of using such a system as a $p$-type or an $n$-type 
semiconductor by fixing the Fermi level in appropriate places.

Before we end, we would like to mention that in the present review we
have addressed some important aspects of quantum transport through some
low-dimensional model quantum systems. Several other important quantum 
phenomena in such meso- and nano-scale systems have also been reported 
in recent reviews~\cite{rev1,rev2,rev3,rev4,rev5,rev6}.

\end{document}